\documentclass[numberedappendix,onecolumn]{openjournal}
\usepackage{amsmath,amssymb,graphicx,natbib,xcolor,bm}
\usepackage{hyperref}
\usepackage[capitalise]{cleveref}
\hypersetup{colorlinks=true,linkcolor=blue,citecolor=blue,urlcolor=blue}
\usepackage{aas_macros}   
\newcommand{\Vsq}{\ensuremath{{|{\cal V}|^2}}}
\newcommand{\sigV}{\ensuremath{\sigma_\Vsq}}
\newcommand{\sigtot}{\ensuremath{\sigma_{\rm tot}}}
\newcommand{\Bmax}{\ensuremath{B_{\rm max}}}

\begin{document}

\title{The information content of optical intensity interferometry}
\shorttitle{Information content of intensity interferometry}

\author{Neal Dalal$^{1}$}
\author{Biwei Dai$^{2}$}
\author{Matthew K. Browning$^{3}$}
\affiliation{$^1$Perimeter Institute for Theoretical Physics, Waterloo, ON, Canada}
\affiliation{$^2$Institute for Advanced Study, Princeton, NJ, USA}
\affiliation{$^3$Physics and Astronomy Department, University of Exeter, Exeter, UK}
\shortauthors{Dalal et al.}

\begin{abstract}
Optical intensity interferometry (II)  can in principle
resolve structure on angular scales $\theta\sim \lambda/B$, 
but it measures only the
squared visibility $\Vsq$, discarding Fourier phase. We quantify what information can actually be
recovered from phase-less $\Vsq$ measurements as a function of the aggregate
signal-to-noise ratio and array configuration, using score-based diffusion models as
non-parametric image priors.  We construct an image reconstruction pipeline, and
validate it by reconstructing a variety of images, including GRMHD simulations of black 
hole accretion disks and actual SDO/HMI continuum images of the Sun, 
from phase-less Fourier amplitudes.
Using simulated data, we map the reconstruction fidelity of Sun-like stellar surfaces 
as a function of noise level, number of $(u,v)$ samples, and maximum baseline $\Bmax$, 
for surface maps spanning a range of activity levels. We find that fidelity depends on 
the data only through the total signal to noise ratio (SNR) of the surface features of 
interest, provided that the $(u,v)$ plane is sampled sufficiently densely. 
These results show that the SNR levels achieved by next-generation intensity 
interferometers observing realistic sources are sufficient to image highly complicated 
stellar surfaces.
\end{abstract}
\keywords{stars: imaging --- techniques: interferometric --- methods: statistical}

\maketitle

\section{Introduction}
\label{sec:intro}

Intensity interferometry \citep{HanburyBrownTwiss1956corr,HanburyBrownTwiss1956sirius}
correlates intensity fluctuations, rather than amplitudes, between separated
telescopes. The technique has been used for decades, for example the measurement of stellar diameters by the Narrabri intensity interferometer half a century ago \citep{HanburyBrown1974}, and modern  efforts using imaging atmospheric Cherenkov telescope arrays \citep{Acciari2020magic,MAGIC2024sii,Zmija2023hess,Saha2025ctao,Archer2025gammaCas}.  
More recently, intensity interferometry has seen a
modern revival driven by fast photon-counting detectors
\citep[e.g.][]{Dravins2012prospects,Dravins2013,Dravins2016,Abeysekara2020,Leopold2025hpd,Mozdzen2025sirius,Guerin2025photoncounting,Tolila2024multiplex,Kulkov2024crosstalk,Kulkov2025hbt,Ingenhuett2026miso,Kaiser2026eso,Schweizer2026eonsii,Izraelevitch2026whiterabbit}. 

Because the intensity correlation is formed after photodetection, the technique is
immune to atmospheric phase errors and does not require optical-path stability,
allowing baselines far beyond the reach
of optical amplitude interferometry, e.g.\ kilometers or more. Given the tremendous
value provided by sub-km optical interferometry \citep[e.g.][]{Roettenbacher2016}, 
access to multi-km baselines could revolutionize stellar astrophysics.
The price of phase immunity, however, is the loss of phase information itself: the observable is
the squared visibility $\Vsq(u,v)$, the squared modulus of the Fourier transform of the
source brightness distribution.  For simple sources, this phase-less
data is sufficient to infer gross properties, such as the overall size of
an object, in the context of parametric models.  More general imaging
of arbitrary sources, however, requires solving a phase-retrieval
problem \citep{Fienup1982}. Under certain conditions, such as vanishing noise and extremely dense sampling of the Fourier plane, phase retrieval is a tractable problem amenable to solution by iterative, deterministic algorithms \citep{Shechtman2015phaseretrieval}, but unfortunately the required conditions for these algorithms to succeed are typically not realized in actual astronomical observations.  These classical methods may be assisted by exploiting the
analyticity of the visibility function (Cauchy--Riemann phase recovery;
\citealt{Nunez2012high,Nunez2012imaging}, demonstrated in the laboratory by \citealt{Dravins2015labdemo}), 
or by measuring high-order correlators to determine closure phases 
\citep{NunezDomiciano2015threetel,Zmija2026thirdorder}, 
but at present phase retrieval in astronomical intensity interferometry remains little-used.
A related approach that avoids some of the pitfalls of classical phase retrieval (see Appendix~\ref{app:hio}) is to replace the iterative solver with a trained neural network \citep{Rai2025gan} to determine point estimates of the image solutions.

A promising alternative to traditional phase retrieval is to recast image reconstruction as a Bayesian inverse problem, where the posterior distribution of the image is inferred given the intensity interferometry data and an appropriate prior. The choice of prior is crucial, as it encodes prior knowledge about the underlying astrophysical source. Recent advances in deep generative models, particularly diffusion models, offer a powerful framework for learning data-driven priors from numerical simulations or analytical models. These models capture complex, high-dimensional probability distributions, allowing us to generate realistic astrophysical images that are consistent with known physical constraints. By leveraging these priors, we can sample the posterior distribution of the image given intensity interferometry observations, enabling robust and high-fidelity reconstructions even in challenging observational conditions.
Modern interferometric imaging in the radio has increasingly 
adopted Bayesian formulations with learned or regularized image priors
\citep{EHT2019,Dia2025iris}.  Score-based diffusion models
\citep{SongErmon2019,HoDDPM2020,SongSDE2021} have been used successfully for astrophysical inverse problems
including strong-lensing source reconstruction \citep{Adam2022posterior}, general
inverse imaging \citep{Feng2023priors}, and radio interferometric imaging
\citep{Dia2025iris,Lai2025closurediffusion}. 

In this paper, we demonstrate the efficacy of a generative diffusion model for intensity interferometric image reconstruction. We apply our method to the problem of resolved imaging of stellar disks, using a simple toy model for disk surface features to generate training sets.  Our results show that the diffusion-based approach not only succeeds in image reconstruction, but it significantly outperforms existing classical methods, particularly in scenarios with incomplete UV-plane coverage or high observational noise. Additionally, we investigate the robustness of our approach when trained on misspecified priors and find that it still recovers the true underlying image structure in the high-SNR regime. These findings highlight the potential of generative models in astrophysical imaging and suggest that data-driven priors could play a key role in the future of intensity interferometry.

The structure of the paper is as follows.  In \S\ref{sec:interferometry}, we briefly review intensity interferometry and phase retrieval.  In \S\ref{sec:methods} we describe details of our reconstruction method, including the training of the diffusion prior and our choice of posterior sampling algorithm. In \S\ref{sec:validation} we demonstrate that this method succeeds in phase retrieval for data sets with high SNR, and further show that prior misspecification can be overcome, illustrating with reconstruction of out-of-distribution examples like SDO images of the Sun. In \S\ref{sec:disks} we apply the framework to reconstruct images of black hole accretion disks from GRMHD simulations with analytical priors, demonstrating that this approach works for different applications with different levels of complexities. In \S\ref{sec:results} we show how the quality of image reconstruction depends on overall SNR and baseline coverage.  We conclude in \S\ref{sec:discussion}.

\section{Review of Intensity Interferometry} \label{sec:interferometry}

In this section, we briefly review some pertinent aspects of intensity interferometry.  For a more detailed discussion, see \citet{Dalal2024} and references therein.

Intensity interferometry relies on the super-Poissonian statistics of light emitted by broadband sources like stars or thermal blackbodies.  The interference of waves at different frequencies produces random ${\cal O}(1)$ fluctuations in the intensity of light, with a coherence time $t_c$ that scales inversely with the bandwidth, $t_c \propto \Delta\nu^{-1}$.  By using sufficiently small bandwidths, the coherence time can be made large enough that it is detectable using modern single photon detectors that provide picosecond timing precision \citep{Korzh2020,becker2022,Fleming2025}, enabling these intensity fluctuations to be detected above Poisson noise. 
Telescopes at different locations observe correlated intensity fluctuations, and the amplitude of the cross-correlation is related to a quantity called the {\em visibility} that measures the Fourier transform of the sky brightness.  More precisely, suppose the specific intensity of light at frequency $\nu$ and sky location $\hat n$ is $I_\nu(\nu, \hat n)$, and suppose that we observe the sky with two identical telescopes separated by baseline vector $\bm{B}$.  We define the visibility as
\begin{equation} \label{eq:visibility}
V(\nu, \bm{B}, \Delta t) = 
\int I_\nu(\nu,\hat n) 
\exp\left(i [\bm{k} \cdot \bm{B} - \omega\Delta t] \right)\, d^2 \hat n ,
\end{equation}
where $\omega=2\pi\nu$ and $\bm{k}=(\omega/c)\hat n$.  We can next define a normalized visibility over bandwidth $\Delta\nu$ as
\begin{equation}\label{eq:normvis}
{\cal V}(\nu_0, \Delta\nu, \bm{B}, \Delta t) = \frac
{\int_{\nu_0-\Delta\nu/2}^{\nu_0+\Delta\nu/2} d\nu\, V(\nu, \bm{B}, \Delta t)}
{\int_{\nu_0-\Delta\nu/2}^{\nu_0+\Delta\nu/2}d\nu \int d^2 \hat n \, I_\nu(\nu,\hat n)}.
\end{equation}
The cross-correlation of intensity fluctuations at two different telescopes observing for time $T$ is given by 
\citep{Purcell1956,HanburyBrownTwiss1957theory}
\begin{eqnarray}
\langle\delta N_i \, \delta N_j\rangle  &=& \frac{\Gamma^2}2 
\int dt_i dt_j\, |{\cal V}(\nu_0, \Delta\nu, \bm{B}, t_i-t_j)|^2 
\nonumber \\
&=& \frac{\Gamma^2 T}2 \int d\Delta t\,
|{\cal V}(\nu_0, \Delta\nu, \bm{B}, \Delta t)|^2,
\label{eq:countscovariance}
\end{eqnarray}
where $\delta N_i$ is the fluctuation in the number of photons observed at telescope $i$ over time interval $T$ centered on time $t_i$, the mean photon rate is $\Gamma = A F_\nu \Delta\nu/(h\nu_0)$, $F_\nu\equiv\int I_\nu d^2\hat n$, $A$ is the collecting area of each telescope, 
$\Delta t = t_i - t_j$ and the $t_i$ and $t_j$ integrals run over time interval $T$.

Eq.\ \eqref{eq:countscovariance} shows how the cross-correlation of photon counts at two different telescopes is related to the squared visibility, $|{\cal V}(\bm{B})|^2$.  Measurements of intensity correlations can therefore be used to measure the squared visibility, which (by Eqs.\ \eqref{eq:visibility} and \eqref{eq:normvis}) is related to the Fourier transform of the sky brightness.  
A useful concrete example to consider is
a broadband source with a smooth emission spectrum, observed over a sufficiently narrow bandwidth  ($\Delta\nu \ll \nu_0$) that we can treat the spectrum as being almost constant, $I_\nu(\nu, \hat n) \approx I_\nu(\nu_0, \hat n)$ across the observed bandwidth.  If the source is centred at location $\hat n_0$ on the sky, it is convenient to measure sky locations $\hat n$ relative to this reference location, writing $\hat n = \hat n_0 + \delta\hat n$, so that $\delta \hat n\cdot \hat n_0 \approx 0$ for $|{\delta \hat n}|\ll 1$.  We can decompose the baseline separation vector $\bm{B}$ into components parallel and perpendicular to $\hat n_0$, writing $\bm{B} = \bm{B}_\perp + (\bm{B}\cdot\hat n_0) \hat n_0$.  Then $\bm{k}\cdot\bm{B} = \bm{k}\cdot\bm{B}_\perp + \omega \bm{B}\cdot\hat n_0 /c$, and so $\bm{k}\cdot\bm{B} - \omega \Delta t = \bm{k}\cdot\bm{B}_\perp - \omega (\Delta t - \bm{B}\cdot\hat n_0 /c)$. Let us redefine $\Delta t$ to absorb the last term, i.e.\ $\Delta t \to \Delta t- \bm{B}\cdot\hat n_0 /c$.
In this limit, the time dependence of $\cal V$ in Eqn.\ \eqref{eq:normvis} factorizes out, as 
\begin{equation}
\label{eq:normvis_approx}
{\cal V} \approx {\rm sinc}\left(\frac{\Delta\omega\,\Delta t}2\right) e^{-i\omega_0 \Delta t} \,
{\cal V}(\nu_0, \bm{B}) 
\end{equation}
where  
\begin{equation} \label{eq:vis}
{\cal V}(\nu_0, \bm{B}) = \frac{V(\nu_0, \bm{B})}{I_{\rm tot}} 
= \frac{\int d^2\hat n\, I(\nu_0,\hat n) e^{2\pi i\bm{B}_\perp\cdot\hat n/\lambda_0}}
{\int d^2 \hat n\, I(\nu_0,\hat n)} ,
\end{equation}
and $\lambda_0 = c/\nu_0$.  Note that in the aperture synthesis literature, $\bm{B}_\perp/\lambda_0$ is conventionally written as the two-component vector $(u,v)$. 

Measurements of intensity correlations can therefore be used to measure the squared visibility $|{\cal V}|^2$ from Eq.\ \eqref{eq:vis}.  The noise on measuring $|{\cal V}|^2$ arises from Poisson noise in the photon counts, as well as timing jitter from the photodetectors.  Following the notation of \cite{Dalal2024}, a convenient way to express the noise on $\Vsq$ is the noise variance $\sigV ^2$, defined via
\begin{equation}\label{eqn:sigV}
\sigV^{-1} = \frac{d\Gamma}{d\nu} 
\left(\frac{T_{\rm obs}}{\sigma_t}\right)^{1/2} n_c^{1/2}\, (128\pi)^{-1/4}.
\end{equation}
Here, $d\Gamma/d\nu = A\,F_\nu/(h \nu_0)$ is the rate of photon
detections per unit frequency bandwidth, $T_{\rm obs}$ is the total
observation time, $\sigma_t$ is the standard deviation of the
photodetector timing jitter, and $n_c$ is the number of independent
frequency channels.  The signal-to-noise ratio of the squared
visibility per baseline then becomes $\Vsq / \sigV$.  
We can define the total combined noise for an array with many baselines as
\begin{equation} \label{eq:sigtot}
  \sigma_{\rm tot} = 
  \left[\sum_i \sigma_{\Vsq,i}^{-2}  \right]^{-1/2} 
  = N_B^{-1/2}\sigV,
\end{equation}
where $N_B$ is the number of independent baselines and the second equality assumes that each baseline has identical noise $\sigV$.  The total signal to noise
ratio for an unresolved (e.g., pointlike) source with $\Vsq=1$ is then ${\rm SNR} = \sigma_{\rm tot}^{-1}$.
This combined SNR is distributed among noisy measurements
on different baselines.  In traditional aperture synthesis in radio interferometry \citep{ThompsonBook}, we map out the visibility at many different baselines $\bm{B}$, thereby mapping out the Fourier transform of the sky.  Using an intensity interferometer array, we can perform a similar measurement, mapping out the absolute value squared of the visibility, $|{\cal V}(\bm{B})|^2$, at many points on the Fourier plane.  In comparison to traditional radio interferometry, we immediately see one crucial drawback: we measure only the modulus of the Fourier transform, and cannot measure the phase of the Fourier transform.  The loss of phase information limits our ability to reconstruct an image of the sky brightness using interferometric measurements.  

However, it has long been known that under certain conditions such as
vanishing measurement noise and extremely dense coverage of the
Fourier plane, it is possible to recover much of the phase information that is
lost to an intensity interferometer.  This procedure is known as phase
retrieval, and algorithms to recover phase information have been known
for decades \citep{Fienup1982}.  The application of classical phase
retrieval to intensity interferometry observations has been somewhat
limited, in part due to the onerous requirements of these methods.  In
this paper, however, we argue that modern Bayesian machine learning
methods, applied to realistic intensity interferometry observations,
can effectively achieve phase retrieval for arbitrary sources.  This
is because modern single-photon detectors allow intensity
interferometers to reach extremely high signal to noise ratios.

As an illustrative example, \citet{Dalal2024} considered an array with
a total collecting area of $A_{\rm tot}\approx 1000\,{\rm m}^2$,
modeled on the proposed LFAST array \citep{LFAST}, for which a prototype array is funded and under construction\footnote{For more details on the prototype LFAST-20x array, see \href{https://www.schmidtsciences.org/schmidt-observatory-system/\#modal-large-fiber-array-spectroscopic-telescope-lfast}{the project website}.}.  As they noted, such
an array, with $\sigma_t=13$ ps and $n_c=5000$ would reach
$\sigtot\approx 10^{-3}$ for an extragalactic AGN at magnitude
$V=12$ in about 6 hours.  Since \sigtot\ scales inversely with source 
brightness, a similar
array would reach $\sigtot\,\approx 10^{-7}$ for a bright nearby star with
magnitude $V=2$.  Therefore, realistic arrays can reach astonishing
SNR for actual stars on the sky in reasonable observing times.  For
this reason, it is interesting to consider what information can be
inferred from data sets that reach these SNR levels. 
The principal component analysis of \citet{Dalal2024} suggests a hierarchy of
different information regimes:

\begin{enumerate}
\item \textbf{Size measurement} ($\sigtot\gtrsim10^{-2}$).
At low SNR, only one or two source
modes are measurable, which in practice inevitably involve the overall
angular size of the source, obtained by fitting the low-order shape of
$\Vsq$ near and below the first null. In this regime, since the size
is effectively the only measurable quantity, then the typical baseline
length (which we will characterize by the longest baseline in the
array, $\Bmax$) should be tuned to the source diameter (the first null of a
uniform disk of angular diameter $\theta_d$ lies at $B\simeq1.22\,\lambda/\theta_d$,
$\approx130$\,m for the Sun observed at distance 10\,pc at wavelength $\lambda= 500\,$nm). This is the regime of the
classical diameter programs \citep{HanburyBrown1974} and of many modern
re-demonstrations \citep[e.g][]{Abeysekara2020}.

\item \textbf{Parametric modeling} ($\sigtot\lesssim10^{-3}$). At higher
  SNR, several to tens of modes become measurable.  This is inadequate
  to reconstruct arbitrary sources, but allows fits to low-dimensional
  parametric source models that describe structure in the source on
  scales smaller than its overall size. To probe this smaller-scale
  structure, it is fruitful to increase $\Bmax$ according to the
  angular scale of the modeled features. Because the
model dimension is far below the number of image pixels, no phase retrieval is
needed: the parametric likelihood on $\Vsq$ suffices. 

\item \textbf{Non-parametric imaging} (\sigtot\  below a feature-dependent
threshold). Enough modes are measurable to reconstruct the image
without relying on a
parametric model --- implicit phase retrieval. The threshold depends strongly on the
angular scale and contrast of the targeted features.  
\end{enumerate}
Previous work has extensively studied the first two regimes.   
The purpose of this paper is to explore non-parametric image reconstruction in detail.  We
will use stellar intensity interferometry as a concrete example, but
the information content of this class of observations generalizes to
other types of targets as well, such as active galactic nuclei (AGN).
In \S\ref{sec:methods} we describe our method for Bayesian 
reconstruction of stellar images,  in \S\ref{sec:validation} we
validate our reconstruction method on simulated stellar datasets with
negligible noise, in \S\ref{sec:disks} we apply it to black hole accretion disks, and then in \S\ref{sec:results} we explore the 
different SNR regimes described in the list above.

\section{Methods}
\label{sec:methods}

Our model of the observed stellar image consists of three ingredients,
ordered from global to local. First, we have a small set of geometric
parameters: the angular size of the disk, the viewing
inclination, position angle, and rotational phase. Second, we assume a
circularly symmetric limb-darkening profile $I(\mu)$,
described by a few coefficients (\S\ref{sec:forward}).
Third, smaller-scale features like spots and faculae that are the actual imaging targets are described by
a pixelized HEALPix map $y(\hat{n})$ of the
intrinsic stellar surface.  We could instead describe both the large-scale and small-scale surface with a single non-parametric map, but this would be less efficient.  Since the disk geometry and limb darkening are smooth, low-dimensional structures that are either known or cheaply inferred as parameters, it would be computationally wasteful to train an extremely flexible generative prior to learn this simple parametric behavior.  Our approach instead is to reserve the learned generative prior for the surface map alone, since that is the one ingredient that genuinely requires a flexible, high-dimensional model.  The renderer composes the three ingredients into a disk image, and the forward model maps that image to phase-less $\Vsq$ data.

All components below are implemented in JAX \citep{jax2018} and are auto-differentiable end-to-end, which aids in sampling the posterior.  The same pipeline is used both to generate training data and to
perform inference. 

\subsection{Stellar surface model and training set}
\label{sec:generator}

\begin{figure}
\centering
\includegraphics[width=\textwidth]{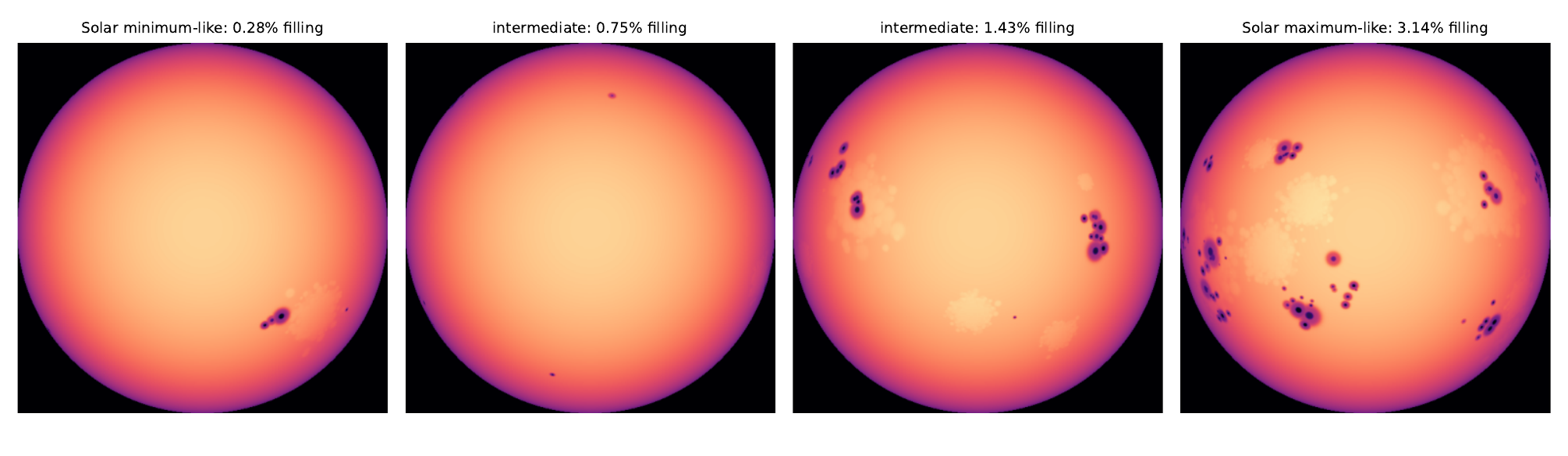}
\caption{Surface-model examples, rendered as observed disk images (orthographic
projection with limb darkening, at the Sun's actual viewing inclination
$i=90^\circ-B_0\simeq85^\circ$, common intensity scale;
dark features are spots, the faint bright patches faculae; panels ordered least to
most active).  As these examples make clear, the training distribution is significantly
unrealistic for actual stellar surfaces.}
\label{fig:trainingset}
\end{figure}

Surfaces are represented as intrinsic (viewing-independent) brightness maps
$y(\hat{n})$ on a HEALPix grid \citep{Gorski2005healpix,Zonca2019healpy} at
$N_{\rm side}=256$ ($786{,}432$ pixels, $\sim$0.23$^\circ$ resolution), with
$y=1$ the quiet photosphere, $y<1$ spots, and $y>1$ faculae.

Surface synthesis builds on the \texttt{spotter} package \citep{spotter}
(v0.0.11), which represents stellar surfaces as HEALPix intensity
maps, and supplies the elementary spot primitive that we adopt: a circular spot of
angular radius $r$ centered at $\hat{n}_0$ has profile
\begin{equation}
w(\hat{n}) = \tfrac{1}{2}\left\{1-\tanh\!\left[\left(d(\hat{n},\hat{n}_0)-r\right)s/2\right]\right\},
\end{equation}
where $d$ is the great-circle distance and $s$ the edge sharpness, so $w\simeq1$
inside the spot and $\simeq0$ outside with a sigmoid edge of width $\sim1/s$.
The edge width is assumed to
scale with the spot size, $1/s=0.125\,r$, capped at the pixel scale.  
We generate the training
population with the following simplified rules:

\begin{itemize}
\item \emph{Spot temperatures and contrasts.} Spot temperature deficits follow the
empirical spot-versus-photosphere temperature relation compiled by
\citet{Berdyugina2005}; brightness contrasts are computed from the Planck function
at the observing wavelength (visible-band contrasts are systematically deeper
than bolometric ones, by up to $\sim$30\% for cool photospheres). Spots are modeled
as an umbra (radius $0.4\,r$, full contrast) inside a penumbra (radius $r$, half
contrast) \citep{Solanki2003}, and overlapping spots merge into a single umbra rather than multiplying toward zero. 
\item \emph{Sizes.} Penumbral radii follow a log-normal distribution
\citep{Bogdan1988} with median $0.02$\,rad ($\approx1.1^\circ$), i.e.\
$\approx$14\,Mm on a solar-radius star. This choice is motivated by our map resolution, not solar realism: most real sunspots are several
times smaller and sub-pixel at $N_{\rm side}=256$ (pixel $\approx$2.8\,Mm), so the
generator populates the star with the \emph{resolvable} tail of the size
distribution. A more faithful solar population would require $N_{\rm side}\ge512$.
\item \emph{Grouping and latitudes.} Spots are placed in active regions of 1--10
spots with overlap avoidance. Region locations follow a three-component mixture.
70\% of regions occur in an equatorial band with latitude $b$ drawn randomly from Gaussians centered on $\pm20^\circ$ with rms of $10^\circ$. 15\% of regions are at polar latitudes with
$b\sim\pm70^\circ$, and the final 15\% of regions are placed isotropically on the surface. 
\item \emph{Filling factor.} Each map draws a target spot filling factor
log-uniformly over $0.1\text{--}5\%$. In comparison, the real Sun sits at
$\sim0.1$--$0.2\%$ in this metric during quiet periods, growing to $\sim 4-5\%$ during highly active periods. Genuinely active stars (filling factors of tens of
percent) are out of scope for this prior.
\item \emph{Faculae.} Facular area is tied to spot area through the
activity-dependent ratio $Q(f)$ of \citet{Shapiro2014}, split between plage
bordering active regions and isolated network patches, with a $+2$--$5\%$
brightening. We neglect center-to-limb facular contrast variation, which is why
bright faculae are commonly found near disk center in our training set, unlike
in real stars like the Sun.
\end{itemize}

The training set comprises $5\times10^4$ such maps; Figure~\ref{fig:trainingset}
shows examples spanning the modeled activity range.
Note that this simple surface model deviates from the real Sun in multiple ways. First, the training set's filling-factor metric
(fraction of $N_p=12 N_{\rm side}^2$ pixels with brightness deficit $>5\%$, i.e.\
including penumbrae and pixelization spread) is  systematically high relative to the real Sun.  Second, a ``quiet'' sample
here displays one or two large resolvable spots (in our low-resolution maps) rather than the scattering of much
smaller spots and pores of the real low-activity Sun. Because smaller features
produce weaker $\Vsq$ signatures at proportionally longer baselines
($B_{\rm feat}\sim\lambda/\theta_{\rm feat}$), the SNR thresholds and optimal
baselines we derive in \S\ref{sec:results} are best-case (lower-bound) values with
respect to the true solar spot-size population: resolving genuinely solar-scale
spots would demand both higher SNR and longer baselines than quoted, in proportion
to the size ratio.  Third, the generator's filling floor of $0.1\%$ means it
never produces a literally spotless star, whereas the real Sun is spotless on most
days near cycle minimum. The training set includes explicitly blank maps as an
augmentation, but the truth population sampled in \S\ref{sec:results} does not. 
We use these unrealistic, low-resolution stellar disk images solely as a convenient
sample for training a diffusion prior inexpensively.  
To plan observations of actual stars, more realistic priors (at
higher resolution) should be used instead.

\subsection{Diffusion prior}
\label{sec:diffusionprior}

Using the training set described above, we construct a prior over stellar surface features.  
Appendix \ref{app:training} provides more detail about the diffusion prior and its training, but we briefly summarize some aspects here.    
We model the prior over surface deficit maps $\delta=1-y$ with a denoising
diffusion probabilistic model
\citep[DDPM;][]{SohlDickstein2015,HoDDPM2020}. Diffusion models learn a
distribution by inverting a gradual noising process: each training map is
corrupted by Gaussian noise of increasing variance over $T$ discrete levels,
and a neural network is trained to undo one noise increment at every level.
This is equivalent to estimating the score $\nabla_\delta \log p_t(\delta)$ of
the noise-convolved distribution at each noise level
\citep{SongErmon2019,SongSDE2021}. New maps are generated by drawing pure Gaussian noise and applying the learned denoising steps in reverse. This reverse pass is a discretization of a stochastic differential equation whose drift is set by the score \citep{SongSDE2021}, and this procedure transforms the Gaussian distribution at $t=T$ into the learned prior at $t=0$. The denoising
network is the sole learned object, and because it provides the score of the
prior at every noise level it can
be combined with a likelihood at matched noise levels to sample a
posterior. \citet{Yang2023survey} reviews diffusion models generally, and
\citet{Daras2024survey} surveys their use for inverse problems of this kind.
We use the recipe of \citet{HoDDPM2020} unchanged, with $T=1000$ steps and a linear variance schedule $\beta_t\in[10^{-4},0.02]$. A training map $\delta$ is noised to level $t$ as $\delta_t=\sqrt{\bar\alpha_t}\,\delta+\sqrt{1-\bar\alpha_t}\,\epsilon$, where $\epsilon\sim\mathcal{N}(0,\mathbb{I})$ and $\bar\alpha_t=\prod_{s\le t}(1-\beta_s)$. The network $\epsilon_\theta(\delta_t,t)$ is trained to recover the injected noise, using their simplified $\epsilon$-prediction loss
\begin{equation}
\mathcal{L} \;=\; \Big\langle
\left\| \epsilon - \epsilon_\theta(\delta_t, t) \right\|^2
\Big\rangle,
\label{eq:epsloss}
\end{equation}
with the average taken over $t$ (drawn uniformly), training maps $\delta$, and noise draws $\epsilon$. Every pixel and every noise level enters
\cref{eq:epsloss} with equal weight, a property that matters for training on
maps as sparse as ours (Appendix~\ref{app:training}).

\subsection{Forward model for \Vsq}
\label{sec:forward}

The forward model converts an intrinsic surface map $y(\hat n)$ from the sphere to the
observable $\Vsq$. 
For this paper, we will assume the surface is static in time.  Actual stars rotate, and
their surface features vary over time (e.g., spots emerge and decay), but we assume here
that integration times are small enough that this time evolution may be neglected, i.e.\ 
high SNR corresponds to arrays with large collecting areas rather than long integration
times.  We did check using images of the Sun (like those in \S\ref{sec:sdo}) that 
short-term variability from processes like granulation does not affect image 
reconstruction at any noise level considered in this work.

The \texttt{spotter} package used by the generator of
\S\ref{sec:generator} provides a reference orthographic renderer for its
surface maps, and we use it as a validation standard, but not in the forward
model itself. The reason is that its internal rotation is performed by NumPy-based
\texttt{healpy} operations rather than JAX operations. The \texttt{spotter} renderer is therefore not automatically differentiable, posing a challenge for
posterior sampling (\S\ref{sec:sampling}) which requires gradients of $\Vsq$ with respect to the surface map, the limb-darkening coefficients, and the stellar angular size. We therefore implement a pure-JAX differentiable orthographic renderer that shares \texttt{spotter}'s map conventions and is validated against its output. The renderer maps an intrinsic map to a disk image for
inclination, position angle, and rotational phase: each image pixel is assigned the bilinearly-interpolated surface value at its sub-pixel $(\theta,\ell)$ coordinate (via HEALPix 4-neighbor interpolation weights), multiplied by a limb-darkening
factor $I(\mu)/I(1)$. We use the general 4-parameter law of \citet{Claret2000},
\begin{equation}
I(\mu)/I(1) \;=\; 1 - \textstyle\sum_{k=1}^{4} a_k\,(1-\mu^{k/2}),
\label{eq:claret}
\end{equation}
which is flexible enough to represent realistic profiles without ringing and
contains the familiar 2-parameter quadratic law,
$I(\mu)/I(1)=1-u_1(1-\mu)-u_2(1-\mu)^2$, as the special case
$(a_1,a_2,a_3,a_4)=(u_1+2u_2,\,0,\,-u_2,\,0)$. 

Just as a prior is required for the surface deficit map $\delta(\hat n)$, we also require a prior on the limb-darkening parameters.  We assume an isotropic Gaussian prior
$a_k \sim \mathcal{N}(a_{0,k},\,0.5^2)$, which is broad enough to permit order-unity
excursions of the profile, and so we truncate the prior range to impose non-negative intensity,
$I(\mu)\ge0$ for all $\mu$. 
The coefficients are inferred jointly with the surface map by Gibbs sampling within the annealed sampler described below in \S\ref{sec:sampling}.  One consequence of peak normalization worth noting is that a uniform change of the disk-integrated flux cancels exactly, so the map-average (DC) mode is constrained only by the prior.

We generate images at $256^2$
resolution, zero-pad by a factor 2, and then Fast Fourier Transform to compute 
the peak-normalized squared modulus $|{\cal V}(u,v)|^2$ on a Nyquist-limited  grid.  In some cases, we make use of the entire Fourier grid, and in other cases we sample baselines to mimic the Fourier coverage provided by telescope arrays.  In the latter case, baseline samples are drawn from an area $B < \Bmax$, then binned
onto the Fourier grid.  When $m$ points land in the same cell, the per-cell noise is $\sigma_c\equiv\sigV/\sqrt{m}$. For array-realistic experiments we use physical
telescope layouts with Earth-rotation synthesis (hour-angle tracks), including a
compact-core design that fills the short-baseline hole of a random array. Gaussian noise of width $\sigV$ per point (or  $\sigma_c$ per cell) is
added to $\Vsq$.

\subsection{Posterior sampling from phase-less data}
\label{sec:sampling}

In \S\ref{sec:diffusionprior} we described how diffusion models provide an elegant way to sample from a learned prior distribution.  With a small modification, the same diffusion model can be used to sample from the posterior distribution, given noisy measurements of data.  The key idea is that the same reverse pass that samples the
prior can also sample the posterior. By Bayes' theorem the score of the
posterior is a sum of two terms,
\begin{equation}
\nabla_\delta \log p(\delta\,|\,\Vsq_{\rm obs}) \;=\;
\nabla_\delta \log p(\delta) \;+\;
\nabla_\delta \log p(\Vsq_{\rm obs}\,|\,\delta) .
\label{eq:postscore}
\end{equation}
The first term is the prior score that the network already provides. The
second is the gradient of the data likelihood, which we obtain by
differentiating the forward model of \S\ref{sec:forward}. Running the reverse
pass with this sum (in place of the prior score alone) pulls each trajectory
toward maps that are both typical of the prior and consistent with the
measured visibilities. This is the approach used for score-based priors in
other astrophysical inverse problems \citep{Adam2022posterior,Dia2025iris}.
Each pass returns one independent sample from the posterior. We call each such
sample a particle, and the ensemble of $K$ independent particles is the
posterior ensemble. 
Where noted below, each particle also draws its own activity level and limb-darkening coefficients.

One complication shapes what follows. Partway through the reverse pass, the map
is still part noise, and the term the sampler needs there is the likelihood of
the data given that noisy map. That likelihood has no closed form for a nonlinear
forward model, and approximating it is the central design problem in diffusion
posterior sampling \citep{Daras2024survey}.  We describe four aspects of our sampler below, and Appendix~\ref{app:sampler}
lists alternative schemes.

\paragraph{Unconditional predictor + Langevin correctors.}
Each posterior sample involves a numerical integration of the reverse-time diffusion SDE from pure noise to a clean map \citep{SongSDE2021}. A predictor step advances this integration from one noise level to the next, and corrector steps then equilibrate the sample at the new level by Langevin dynamics. In our sampler the likelihood enters only through the correctors. At each noise level we take $n_{\rm corr}=16$ corrector steps, with step sizes set by the signal-to-noise rule of \citet{SongSDE2021}. The predictor uses the prior score alone. This scheme is a corrector-only variant of the predictor--corrector framework of \citet{SongSDE2021}, and at the predictor level it is equivalent to annealed Langevin dynamics \citep{SongErmon2019}. The split keeps the sampler stable at high data SNR, where the likelihood gradient is stiff. The corrector step size self-scales with the score, but the predictor step size is fixed by the noise schedule, so a stiff likelihood gradient in the predictor is unregulated. In early experiments we found that placing the likelihood in the predictor destabilized the reverse pass at high SNR. The self-scaled correctors instead absorb the stiff gradient in many small equilibrating steps.  We found this annealed reverse-SDE sampler to be stable even at the highest data SNR considered here.

\paragraph{Jacobian-aware convolved-likelihood annealing (CLA).} 
At every noise level $t$, the correctors need
$p(\Vsq_{\rm obs}\,|\,\delta_t)$, the likelihood of the data given the noisy
map, which is not the measurement likelihood. \citet{Adam2022posterior} obtain
it by approximating $p(\delta_0\,|\,\delta_t)$ as a Gaussian and propagating
that Gaussian through the forward operator. Its mean is the denoised estimate
of the map, $\hat\delta_0(\delta_t) = \bigl(\delta_t -
\sqrt{1-\bar\alpha_t}\,\epsilon_\theta(\delta_t, t)\bigr)/\sqrt{\bar\alpha_t}$,
which is Tweedie's formula \citep{Efron2011tweedie}, and its variance is
$\gamma_t^2\equiv(1-\bar\alpha_t)/\bar\alpha_t$. The symbols $\delta_t$,
$\bar\alpha_t$, and $\epsilon_\theta$ are defined in \S\ref{sec:diffusionprior}.
The likelihood is therefore evaluated at $\hat\delta_0$ rather than at
$\delta_t$, and its variance is inflated above the measured value by an amount
that anneals to zero as $t\to0$. For a linear forward operator $A$ the inflated
covariance is available in closed form, $\sigV^2+\gamma_t^2 AA^{\!\top}$, and
this construction was validated on interferometric imaging by
\citet{Dia2025iris}. The inflation keeps the strong data term from
destabilizing the trajectory at high $t$ while the sample is still mostly noise. One difference in intensity interferometry is that our
phase-less $\Vsq$ operator is quadratic in the image, so no closed form for the covariance is
available. Additionally, because \Vsq\ varies
by orders of magnitude between short and long baselines, a single scalar
inflation cannot temper every Fourier cell appropriately at high SNR. We
therefore extend the construction by linearizing the forward model about the
current denoised estimate and retaining only the diagonal of $JJ^{\!\top}$,
which gives each cell its own annealed variance,
\begin{equation}
\sigma^2_{{\rm eff},c}(t) \;=\; \sigma_c^2 \;+\; \gamma_t^2\,
{\rm diag}(JJ^{\!\top})_c ,
\label{eq:cla}
\end{equation}
where $J=\partial\Vsq/\partial\delta$ is evaluated at $\hat\delta_0$ and
$\sigma_c$ is the per-cell measurement noise of \S\ref{sec:forward}. Each cell
is inflated according to its own forward sensitivity. Well-measured short baselines stay stable early in the anneal, while long baselines, where $\mathrm{diag}(JJ^{!\top})_c$ is small, sharpen as soon as the data allow. Non-uniform data noise, such as
$\sigma_c$ set by coverage redundancy, enters the same machinery naturally. We
estimate the diagonal with stochastic probes \citep{Hutchinson1989},
${\rm diag}(JJ^{\!\top}) = \big\langle(Jv)^2\big\rangle_v$ over random $\pm1$
vectors $v$. Each product $Jv$ is computed by forward-mode automatic
differentiation. It costs one forward-model evaluation, a render and a Fourier
transform, and no network pass. Four probes per noise level suffice, recomputed
at that level's current $\hat\delta_0$ and held through its corrector steps.
Reconstruction quality with \cref{eq:cla} improves monotonically down to
$\sigV=10^{-8}$, where per-sample fidelity approaches 1.00.

\paragraph{Static thresholding.} At every reverse step the denoised estimate
$\hat\delta_0$ is clipped to the physically allowed deficit range
$-0.15 \leq \delta\le1$. This range requires the surface brightness to be
non-negative and the facular brightening to be modest. The noise prediction
is then recomputed to be consistent with the clipped estimate, by inverting
Tweedie's formula:
$\epsilon \leftarrow (\delta_t - \sqrt{\bar\alpha_t}\,\hat\delta_0)/\sqrt{1-\bar\alpha_t}$.
The trajectory therefore can never wander onto unphysical surfaces. Without
this projection, network errors early in the anneal can push the state out of
range and then compound. The risk is greatest early because the map is mostly
noise there and the denoiser is least accurate. The technique is standard in
the diffusion literature under the name static thresholding
\citep{HoDDPM2020,Saharia2022imagen}. It plays the same role as a positivity
constraint in classical image deconvolution. The prior information is imposed
exactly and at no cost, because the true map satisfies it by construction.

\paragraph{Joint limb-darkening sampling.}
The limb-darkening coefficients of \cref{eq:claret} are Gibbs-sampled
alongside the map. The rendered image is linear in the Claret basis, so the
posterior over the four coefficients, conditioned on the current denoised
map, is linearized Gaussian. One conjugate draw per annealing level then
costs only a $4\times4$ solve. Draws that violate the $I(\mu)\ge0$ constraint
are rejected and redrawn within the Gaussian conditional, making this a
truncated-Gaussian Gibbs step. Each particle carries independent
coefficients, so the ensemble marginalizes over the limb-darkening
uncertainty. We will see in \S\ref{sec:results} that at high SNR the data
constrain the limb darkening so tightly that sampling the coefficients and
holding them fixed give indistinguishable results. For that reason, the
calculations near the end of \S\ref{sec:results} hold the Claret coefficients
fixed.

\subsection{Fidelity metrics}
\label{sec:metrics}

All map metrics are computed on rendered disk images. Let $I(\mathbf{x})$ be a
rendered image of the disk, $D_0(\mathbf{x})$ the rendered image of the unspotted star, determined by the diameter and limb darkening parameters, and 
$r \equiv I - D_0$ the spot residual, evaluated over the visible-disk pixels
$\mathcal{D} = \{\mathbf{x}: D_0(\mathbf{x})>0\}$. Our image generator creates truth images, i.e.\ noiseless realizations of $I(\mathbf{x})$, $D_0(\mathbf{x})$, and $r=I-D_0$, that we label as $r_{\rm true}$.  The sampler generates a different realization, using the noisy \Vsq\ data, that we label as $r_{\rm rec}$.  Our fidelity metrics aim to quantify the agreement or disagreement between $r_{\rm true}$ and $r_{\rm rec}$.  Because $\Vsq$ is invariant under certain trivial transformations of the data, such as $\pi$ rotations or uniform translations, we maximize our fidelity metrics over these trivial transformations.  Let us denote this maximization as $g(r_{\rm rec})$.

The most obvious (but somewhat less useful) metric is the pixel-wise cross-correlation coefficient between $I_{\rm true}$ and $g(I_{\rm rec})$.  This correlation is dominated by the limb-darkened disk and therefore carries a high floor (any reconstruction with approximately the right
size and limb darkening scores $\gtrsim0.95$), so its information is
concentrated in the approach to unity.  We mainly use this to assess the reconstruction of disk diameters and limb darkening profiles.  But because the correlation of $I$ is less sensitive to the fidelity of small surface features, 
our primary metric is the pixel-wise cross-correlation coefficient of the residuals $r_{\rm true}$ and $g(r_{\rm rec})$, 
\begin{equation}
\texttt{spot\_corr} =
\frac{\sum_{\mathcal{D}} r_{\rm true}\, g(r_{\rm rec})}
     {\sqrt{\sum_{\mathcal{D}} r_{\rm true}^2\;\sum_{\mathcal{D}} g(r_{\rm rec})^2}}.
\label{eq:spotcorr}
\end{equation}

Additionally, we also define fidelity metrics to distinguish between gross-structure of the surface features and fine (small-scale) structure.  
We split the residual $r$ with a moving-average filter
$L[\cdot]$ of width 12 pixels ($\approx0.09$ stellar radii at our image scale):
the \emph{envelope} metric applies \Cref{eq:spotcorr} to $L[r_{\rm true}]$ and
$L[g(r_{\rm rec})]$, and the \emph{fine-structure} metric to the complements
$r-L[g(r)]$. The envelope metric tracks the low-order flux
distribution (active-region placement), while the fine structure metric tracks the
individual-spot detail that is probed by the longest baselines.

Every run also reports 
$\chi^2$ against its statistical floor,
$\mathrm{misfit} = \sum_c (\Vsq_{{\rm model},c}-\Vsq_{{\rm obs},c})^2/\sigma_c^2
\,/\, N_{\rm cells}$, and a
``roughness ratio'' comparing the high-frequency map power of posterior samples
against prior samples, aimed at diagnosing speckle structure which data-fit metrics alone cannot
detect.

\section{Reconstruction of stellar disks} \label{sec:validation}

In this section, we apply the pipeline described above in \S\ref{sec:methods} to image reconstruction of stellar disks.

\subsection{Comparison to classical phase retrieval algorithms}
\label{sec:hio}

\begin{figure}
\centering
\includegraphics[width=0.9\textwidth]{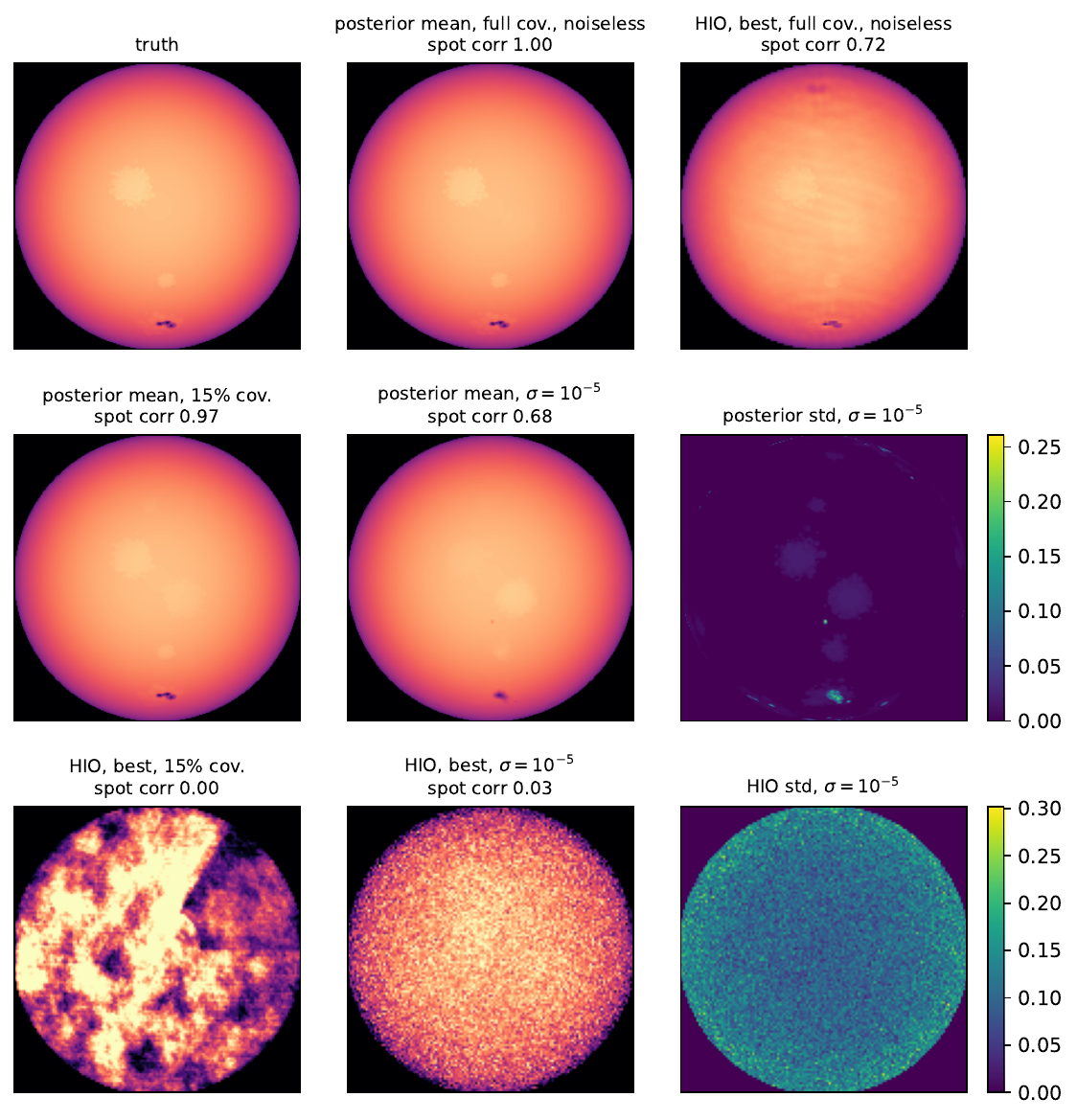}
\caption{Comparison between image reconstruction using diffusion priors
and classical phase retrieval. 
\emph{Top row:} 
The true image to be reconstructed (left), posterior mean over 8
samples (middle), the best of 16 HIO solutions (right), using noiseless \Vsq\ over 
a doubly oversampled Fourier plane, as required for classical phase retrieval. 
In this regime, the diffusion model recovers the image with perfect fidelity,
while the HIO reconstruction exhibits significant artifacts.
\emph{Middle row (posterior):} the posterior mean at
$15\%$ Fourier coverage (noiseless, left); mean at
$\sigma=10^{-5}$ full coverage (middle); and per-pixel standard deviation 
(right). \emph{Bottom row (HIO):} Same \Vsq\ data as the middle row, but
using classical phase retrieval.  Instead of posterior means, the best 
(lowest data residual) reconstructions are shown.  See
Appendix~\ref{app:hio} for the full methodology and mechanism.}
\label{fig:hio}
\end{figure}

We first compare our diffusion-based image reconstruction method with 
classical iterative phase retrieval on identical data. 
Classical phase retrieval reconstructs images using very few (and weak) assumptions about the structure of the image, such as non-negative pixel values.  This demonstrates that even a minimal amount of prior information about the structure of images can enormously improve our ability to recover Fourier phases.  However, significant challenges to classical phase retrieval have been known for many years, such as its fragility to noise in \Vsq, or its rather extreme requirements on sampling of the Fourier plane.

We implemented Fienup's hybrid input-output (HIO)
algorithm \citep{Fienup1982} and applied it to the same phase-less \Vsq\ data
that our sampler faces; see Appendix \ref{app:hio} for more detail.
\Cref{fig:hio} shows the comparison for an example true image (top-left panel), 
with 3 example \Vsq\ data sets.  On noiseless data that doubly oversamples
the Fourier plane (top row), HIO reaches $\texttt{spot\_corr}=0.72$, settling
on a reconstruction that is a stable mixture of the true image and 
its $\pi$ rotation, along with other visibly apparent defects.   We believe 
this solution had reached convergence (see discussion in Appendix~\ref{app:hio}), 
although it is possible that with many orders of magnitude further iterations, 
the method could have found a higher-fidelity solution.  In comparison, for 
the same dataset the diffusion-model posterior samples all achieved nearly
perfect fidelity.  Reducing coverage to $15\%$ of Fourier cells
(left column) leaves HIO returning structureless
noise ($\texttt{spot\_corr}=0.00$) despite an excellent data residual in \Vsq, 
while the diffusion-prior posterior mean on the identical measured cells 
recovers the map with $\texttt{spot\_corr}=0.97$. If we instead add noise at 
$\sigV=10^{-5}$  (middle column), the threshold noise level where structure 
is expected to become marginally detectable using the metrics discussed in 
Appendix \ref{app:Wiener},  HIO again collapses to noise
($\texttt{spot\_corr}=0.03$), while the posterior recovers real structure
($\texttt{spot\_corr}=0.68$). In the bottom two rows, the right column shows 
the per-pixel standard deviation of the ensemble behind each method's point 
estimate at this noise level (posterior samples above, HIO restarts below). 
The contrast is
qualitative, not just quantitative: the posterior's uncertainty concentrates
on the true features and their $180^\circ$-rotated mirror images, an
informative signature of the actual ambiguity in the data, whereas HIO's
restart-to-restart spread is uniform noise texture over the whole disk,
carrying no information about where the reconstruction might be right or
wrong. These regimes isolate what a learned prior adds: coherence across the
degenerate modes at high SNR, a likelihood that fits genuine signal 
over noise, and a prior that fills unmeasured Fourier cells with
plausible structure instead of leaving them as free parameters that cause stagnation in the (classical) reconstruction algorithm.
These examples illustrate that when noise increases or Fourier sampling decreases, a key difference between posterior sampling and classical phase 
retrieval is that image reconstruction fidelity degrades
gracefully, not catastrophically.  We will see this behaviour more clearly below in \S\ref{sec:results}.

\subsection{Reconstruction of out-of-distribution images} \label{sec:ood}

\begin{figure}
\centering
\includegraphics[width=\textwidth]{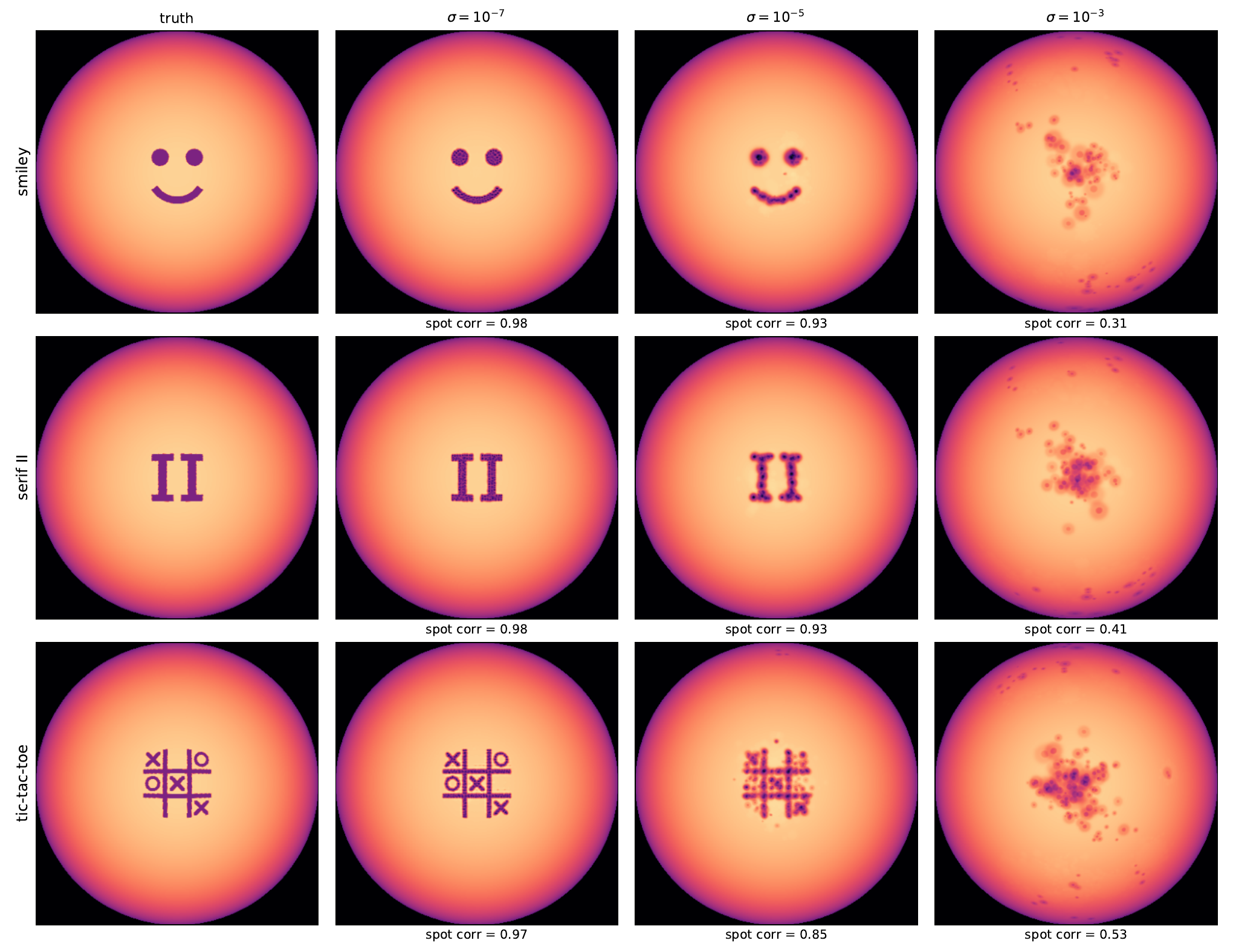}
\caption{Out-of-distribution reconstruction suite. Rows: three hand-drawn truths
the training generator cannot produce. Columns: the truth, then the posterior-mean reconstruction (averaged over 4 samples)
from phase-less $\Vsq$ at full Fourier coverage and varying \sigV. The feature correlation of each mean against
the truth is given beneath each panel. }
\label{fig:ood}
\end{figure}

One obvious concern with the example shown in \cref{fig:hio} is that it shows reconstruction of an image that is typical
of the diffusion prior, and so it may paint an overly rosy picture of how well this method would fare
when applied to more realistic data.  Because our prior model relies on
a training set generated using an extremely simplified model for
stellar surface features, our prior is necessarily misspecified for
actual stars, and so it is important to probe prior misspecification. 
For this reason, we next consider example truth images that the generator cannot produce,
i.e. out-of-distribution examples.  
\cref{fig:ood} shows reconstruction of three deliberately artificial
truths and three noise levels, using the full production pipeline, applied to mock data with full Fourier coverage at varying noise levels.  As the figure illustrates, at very high SNR, we again obtain essentially perfect image reconstruction from phase-less data.  Since these examples are out-of-distribution, this illustrates that our method can recover the truth even when the prior is somewhat misspecified.  To be clear, each truth image here uses ordinary dark-spot amplitudes and an in-distribution filling factor, but the arrangement is not represented in the training set.  As with our in-distribution tests, as we lower the SNR, image reconstruction degrades gracefully, essentially reverting to the prior.  The out-of-distribution features recovered perfectly at high SNR smoothly transition to more prior-typical features at lower SNR.  Therefore, as expected in a Bayesian analysis, although prior misspecification can bias inference, that bias can be overcome with sufficiently high SNR (as long as the prior does not literally vanish at the truth). 

\subsection{Real solar data} \label{sec:sdo}

\begin{figure}
\centering
\includegraphics[width=\textwidth]{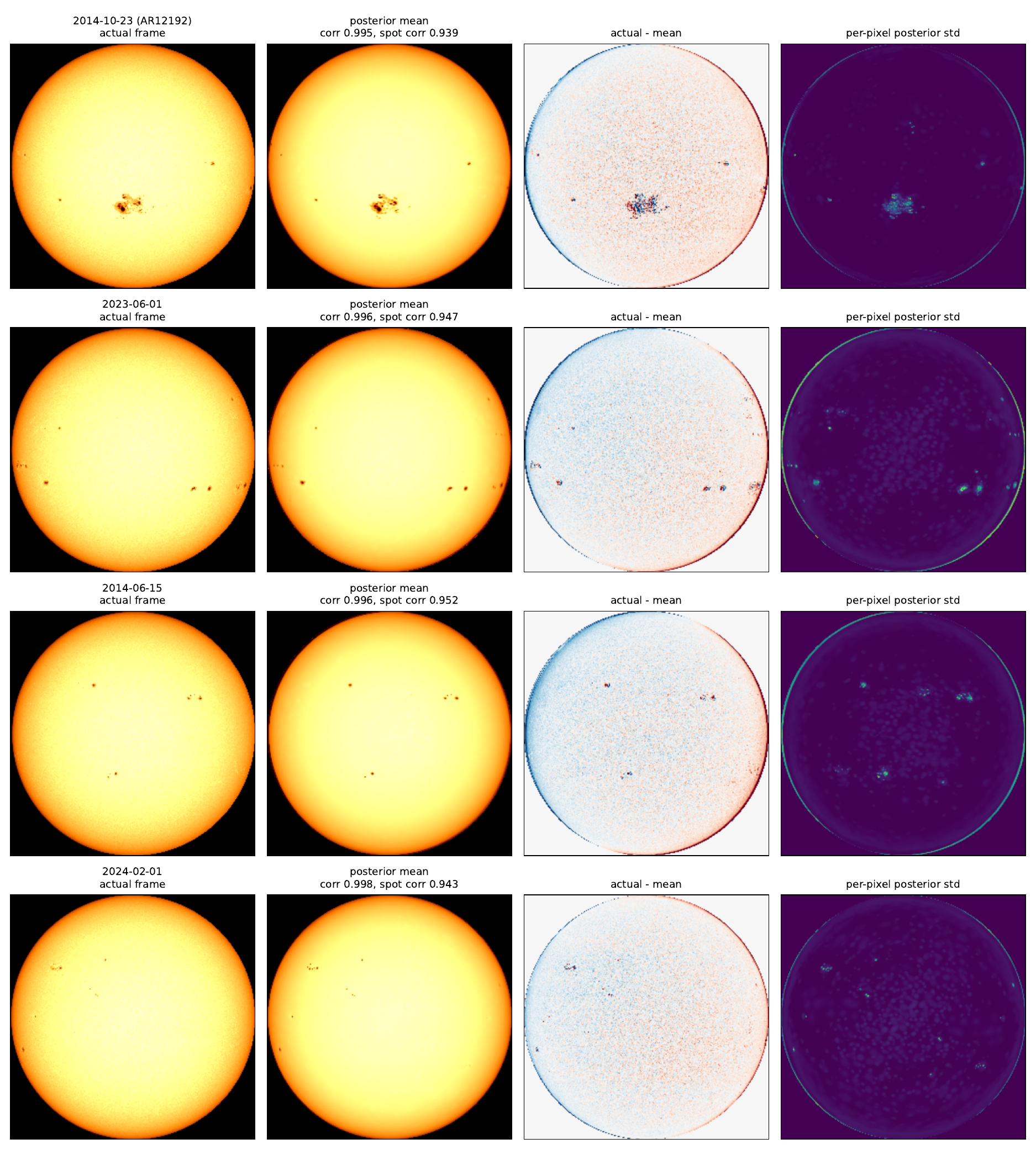}
\caption{Reconstruction of actual observed SDO/HMI continuum images
(derived from 6173\,\AA\ exposures) on various dates.  Each row corresponds
to a different SDO/HMI frame, with columns from left to right showing: 
actual image, posterior mean, residual mean, and scatter among posterior samples.  The posterior scatter is near zero over the quiet disk and
concentrates on the active regions and a thin limb ring, while the quiet-disk
residuals are dominated by granulation-scale texture below the pipeline
resolution, which is structure the surface model does not represent. The faint
east--west residual asymmetry is present in the real data; see main text for more detail.
This figure uses a different color map from the simulated disks elsewhere in the paper, as a visual reminder that these are real solar data.
}
\label{fig:sdo}
\end{figure}

The most important prior misspecification test we can perform, of
course, is to apply our method to resolved images of real stars.  At
present, there is only one star for which high-resolution images are
available, namely our own Sun.  As noted above in \S\ref{sec:methods},
our training data are generated at low resolution to minimize
computational expense, which means that real solar surface features
are highly out-of-distribution for our trained diffusion prior.  
To check the effect of this form of misspecification, 
we reconstruct observed data of the actual Sun, from 
Solar Dynamics Observatory (SDO) \citep{Pesnell2012sdo,Scherrer2012hmi} 
continuum images taken by the Helioseismic and Magnetic Imager (HMI).
Each frame is resampled to the pipeline's pixel scale
($256^2$; the only processing applied), and the phase-less $\Vsq$ data 
are the Fourier amplitudes of that observed image. Nothing else is 
assumed or removed: the limb-darkening profile is inferred jointly from 
the $\Vsq$ data.  As described in \S\ref{sec:sampling}, we model limb 
darkening as a four-parameter Claret law with a broad prior centered on a
uniform disk, constrained only to non-negative intensity. The Sun's
surface is reconstructed with the diffusion prior, which has never seen solar data. 
In addition to the resolution mismatch, the training population's spot contrasts were evaluated at 500 nm, whereas the data are at 6173 \AA, where the same temperature deficit yields a 10–15 percent shallower brightness contrast. The observed HMI contrasts are shallower still, since the standard continuum product is not corrected for instrumental stray light, which fills in umbrae at the tens-of-percent level \citep{Norton2026}.

The reconstruction reproduces the observed images to a pixel correlation of
$0.995-0.998$ (posterior mean). The per-pixel posterior standard deviation 
concentrates on visible spot features and a thin ring at the limb; 
the residual over the quiet disk is noise-like, 
dominated by granulation-scale texture below the pipeline resolution. 
The only coherent residual systematic is a faint east–west asymmetry. One possible origin is the HMI observables algorithm itself: the HMI ``continuum'' is not measured directly but reconstructed from Doppler-shift, line-width, and line-depth estimates of the Fe~{\sc i} 6173\,\AA\ line under a Gaussian-profile assumption \citep{Couvidat2016hmi}, so the retrieved intensity inherits a dependence on line-of-sight velocity.  Across the disk, this is related to the solar rotation. Our forward model applies a circularly symmetric limb-darkening law and so cannot represent any azimuthal asymmetry by construction, leaving such a signal in the residual. We also note that a residual sub-pixel misregistration, to which phase-less data are blind, could produce a qualitatively similar limb-concentrated feature. In either case the amplitude is small and does not affect our conclusions.
This reconstruction also provides an
independent, non-circular cross-check of the jointly inferred limb darkening.
The $I(\mu)$ profile recovered purely from the $\Vsq$ data matches the profile
measured directly from the image (a fit the $\Vsq$ inference never saw) to
$|\Delta I(\mu)|\le0.006$ for $\mu>0.3$, with posterior width $\sim0.001$, and the
two separate only at the extreme limb ($\mu\lesssim0.2$), where $\Vsq$ has
little leverage and the image-domain fit is itself an extrapolation.

The examples shown in \cref{fig:ood,fig:sdo} demonstrate that prior misspecification does not necessarily impede image reconstruction, when the SNR is sufficiently large.  This should make sense from the Bayesian perspective: the posterior is the product of the prior and the likelihood, so a mis-specified prior can bias the posterior when the SNR is low, but at high SNR the likelihood can overpower a biased prior.  Appendix \ref{app:Wiener} discusses the SNR required to overcome various kinds of prior misspecification. \cref{fig:ood} also illustrates how the posterior reverts to the prior at moderate-to-low SNR.  In \S\ref{sec:results}, we study in more detail how image reconstruction depends on the SNR level.

\section{Reconstruction of black hole accretion disks}
\label{sec:disks}

The reconstructions of \S\ref{sec:validation} all target stellar surfaces, whose features are sparse, compact, and high-contrast against a smooth limb-darkened disk of exactly known support. That morphology is favorable in specific ways, and it is reasonable to ask how much of our result depends on it. We therefore apply the same pipeline to a source class with none of those properties: black hole accretion disks, whose emission is continuous, spans orders of magnitude in surface brightness, has no sharp boundary, and carries structure on all scales at once.

This target class has its own compelling science case.
\citet{Dalal2024} showed that realistic intensity interferometer arrays reach sufficient sensitivity to constrain the properties of AGN accretion disks, and analyzed what could be inferred using parametric models, including understanding accretion disk physics and measuring Hubble constant. Similarly, \citet{MoranVanTilburg2026microlens} recently explored intensity interferometry of microlensed quasars, and the QUASAR project \citep{Walter2023accretion} has taken the first steps towards resolved interferometry of bright AGN disks.  While parametric models of accretion disks are a useful benchmark, real disks are not fully described by a smooth parametric model. Spiral density waves, turbulent structure, warps, and vertically extended flows are all outside a radial profile family by construction, and they are among the features that observations would most usefully constrain. They also threaten the parametric measurements themselves: \citet{Dalal2024} note that small-scale fluctuations contribute power at long baselines and partially fill in the visibility nulls, and it is precisely the null structure from which the inner truncation radius is inferred. This is the same distinction drawn in \S\ref{sec:results} between the parametric profile reconstruction of Fig.~\ref{fig:parametric_disk} and non-parametric imaging: the two require different SNR, and answer different questions. We therefore ask whether the reconstruction pipeline developed above can recover disk images non-parametrically, and whether it can recognize a source that the parametric family cannot describe at all.

\paragraph{Training model.}  We build an analytic family from three ingredients. 
First, a \citet{Shakura1973} temperature profile, generalized to an adjustable slope,
\begin{equation}
T(R) \;\propto\; \left[ (R_0/R)^{p} \left( 1 - \sqrt{R_{\rm in}/R} \right) \right]^{1/4} ,
\label{eq:sstemp}
\end{equation}
which recovers the standard profile at $p=3$, with monochromatic intensity $I \propto \left[ \exp(h\nu/kT) - 1 \right]^{-1}$ truncated to zero inside $R_{\rm in}$. 
Second, a multiplicative logarithmic spiral pattern,
\begin{equation}
1 + A \left( 2 \left| \cos \left[ \tfrac{N}{2} \left( \phi - \phi_0 - \ln R / \tan\alpha \right) \right] \right|^{k} - 1 \right) ,
\label{eq:spiral}
\end{equation}
with $N$ arms, pitch angle $\alpha$, contrast $A$, and a sharpness exponent $k$ that interpolates between broad and narrow arms. Third, a multiplicative perturbation to the local temperature, drawn as a Gaussian random field with power-law spectrum $P(q) \propto q^{n}$ and amplitude $\sigma_{\rm turb}$, representing turbulent structure. All parameters are drawn from broad uniform priors, and training images are generated on the fly. The injected turbulent field serves a second purpose beyond realism: it prevents the training distribution from being over-smooth, ensuring that the prior places support on maps with non-zero power at small scales probed by the longest baselines. As in \S\ref{sec:generator}, this model is a convenient way to obtain a broad, physically motivated distribution of disk-like images, and is not meant to represent realistic accretion physics.

\paragraph{Prior and likelihood.}  The prior is trained with the EDM recipe of \citet{Karras2022edm}. Writing the diffused image as $x_t = x_0 + t\epsilon$, the Fourier modulus of the noisy image has, conditional on $x_0$, exactly
\begin{equation}
\big\langle |\hat{x}_{t,\bm{u}}|^2 \big\rangle = |\hat{x}_{0,\bm{u}}|^2 + d\,t^2 ,
\qquad
\mathrm{Var}\big[ |\hat{x}_{t,\bm{u}}|^2 \big] = d^2 t^4 + 2 d\, t^2 |\hat{x}_{0,\bm{u}}|^2 ,
\label{eq:quadmoments}
\end{equation}
with $d$ the number of pixels. Moment-matching a Gaussian to this distribution gives the tempered likelihood
\begin{equation}
\log p \;\simeq\; - \sum_{\bm{u}}
\frac{ \left( |\hat{x}_{t,\bm{u}}|^2 - |\mathcal{V}|^2_{{\rm obs},\bm{u}} - d t^2 \right)^2 }
     { 2 \left( \sigma^2_{|\mathcal{V}|^2} + d^2 t^4 + 2 d t^2 |\mathcal{V}|^2_{{\rm obs},\bm{u}} \right) } ,
\label{eq:quadlike}
\end{equation}
which replaces the Jacobian-probed variance inflation of \S\ref{sec:sampling} with the exact second moment of the quadratic operator, evaluated at $x_t$ rather than at the denoised estimate. 

\begin{figure*}
\centering
\includegraphics[width=\textwidth]{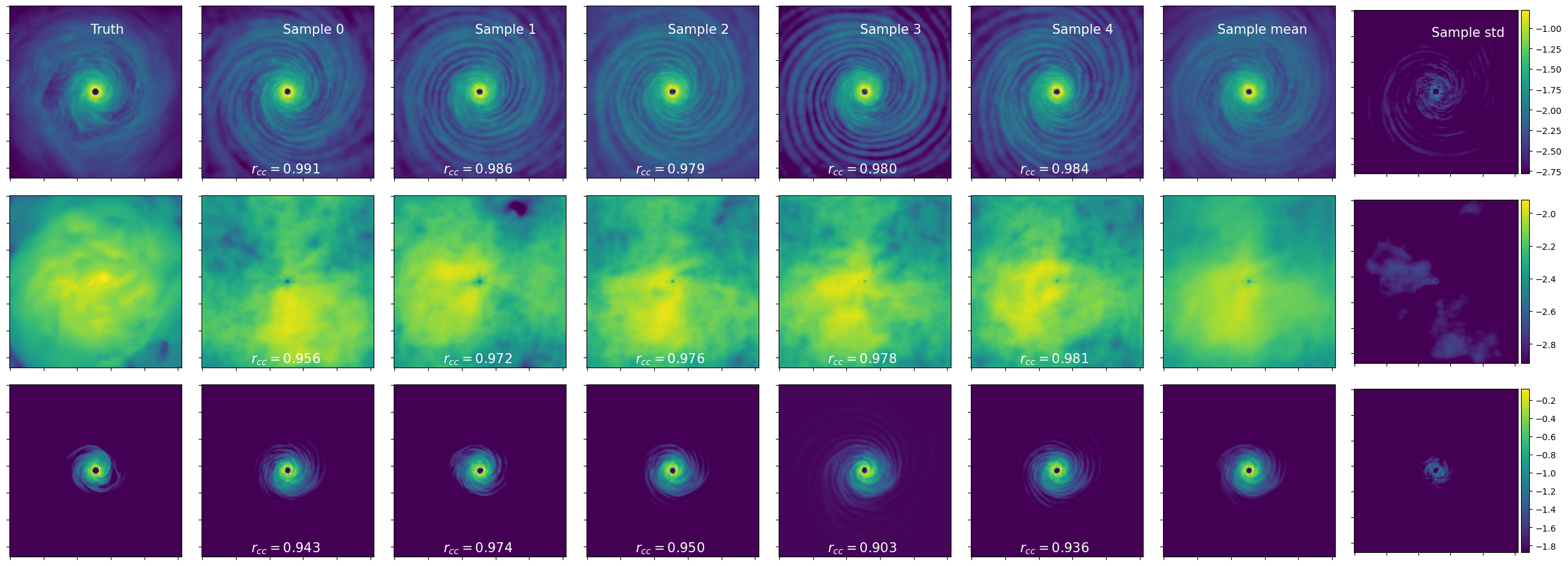}
\caption{Reconstruction of radiation-GRMHD accretion-disk images 
\citep{Zhang2025radiationI,Zhang2026radiationII,Zhang2026radiationIII} from phase-less $|\mathcal{V}|^2$ at full Fourier coverage and $\sigma_{|\mathcal{V}|^2} = 10^{-7}$, using a diffusion prior trained on the analytic disk family of \S\ref{sec:disks}. Rows correspond to three different simulation models and accretion rates; in the middle row the disk is not directly visible. Columns from left to right: truth, five independent posterior samples labelled with their pixel correlation $r_{cc}$ against the truth, the sample mean, and the per-pixel standard deviation among samples. All panels are logarithmic in surface brightness. Note the small dark point at the centre of every reconstruction, absent from the middle-row truth: it is the prior's inner disk truncation at $R_{\rm in}$, and appears identically in all samples.}
\label{fig:disks}
\end{figure*}

\paragraph{Results.}  Figure~\ref{fig:disks} shows reconstruction of three radiation-GRMHD accretion-disk images from \citet{Zhang2025radiationI,Zhang2026radiationII,Zhang2026radiationIII}, corresponding to different models and accretion rates. These images are out-of-distribution in the sense of \S\ref{sec:validation}: the analytic generator cannot produce them. Since this is just a proof-of-concept experiment, we assume a high SNR limit, with full Fourier coverage and $\sigma_{|\mathcal{V}|^2} = 10^{-7}$. Pixel correlations of individual samples against the truth are $0.90$--$0.99$ across all three.

Two features are worth noting in more detail. In the middle row the disk itself is not directly visible, but is instead obscured by foreground gas. The image shows a diffuse brightness distribution with no thin-disk morphology and no bright inner region, and the reconstructions correctly return diffuse, structureless maps rather than forcing the prior's preferred spiral disk. The data override the prior, as expected at this SNR from the argument of Appendix~\ref{app:Wiener}. However, every reconstruction places a small dark point at the image center, including the middle row where the truth has none. This is the prior's inner truncation at $R_{\rm in}$, imprinted at the resolution limit. Because it appears identically in every sample it carries no ensemble variance, and so it would not be flagged by the diversity of the posterior: a feature that the prior imposes universally is precisely the kind that per-pixel scatter cannot diagnose. Taken together, this example shows the pipeline transferring to a morphologically unrelated source class, and the two behaviors are discussed and explained in Appendix~\ref{app:Wiener}.

\section{Information as a function of SNR} \label{sec:results}

\begin{figure*}
\includegraphics[width=\textwidth]{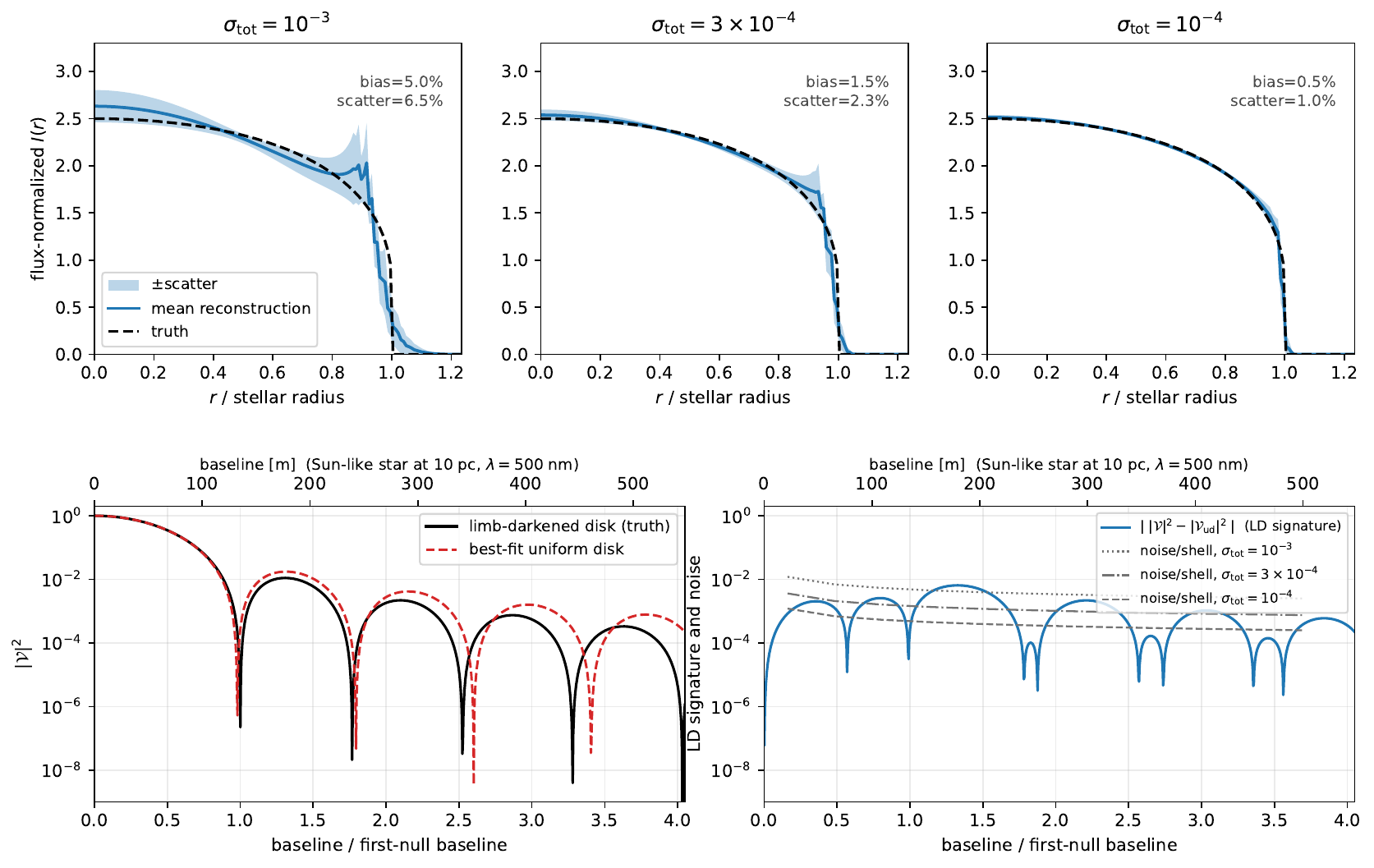}
\caption{Reconstruction with parametric models. \emph{Top row:} the circularly
averaged surface brightness profile $I(r)$, described by a 5-parameter model
consisting of stellar size and the 4-parameter \citeauthor{Claret2000} law. The
true $I(r)$ is indicated by the black dashed curve. Each panel shows a
different \sigtot, along with the mean and scatter of the reconstructed
profiles (blue line and shaded area). \emph{Bottom left:} $\Vsq$ of the true
limb-darkened disk and of the best-fitting uniform disk (size re-optimized):
the two agree through the first null --- the size measurement --- and differ
beyond it in null positions and lobe amplitudes. \emph{Bottom right:} the
limb-darkening signature (the absolute difference of the two curves at left),
together with the per-shell noise levels of the three cases above. The profile
information resides in the lobes beyond the first null, at amplitude
$\sim\!10^{-3}$--$10^{-2}$, and the top-row transition tracks where the noise
crosses this signature: the profile requires $\sigtot\lesssim10^{-3}$ because
that is its signature amplitude --- the same signature-exceeds-noise rule as
the spot features of \cref{fig:retrieval_example}.
}
\label{fig:parametric_disk}
\end{figure*}

In  \S\ref{sec:validation} and \S\ref{sec:disks}, we demonstrated image reconstruction and phase retrieval at high SNR, and we showed examples of how reconstruction degrades as the SNR is lowered.  In this section, we consider more systematically what information about an image may be recovered from phaseless data as a function of signal to noise ratio. Previous work has studied what may be learned from intensity interferometry in the low to moderate SNR regimes, when sizes may be determined and when parametric models may be constrained \citep[e.g.][]{Dravins2016,Bojer2022,Guerin2025photoncounting}.  For example, simple Fisher matrix analysis \citep{Dalal2024} finds that multi-parameter models may be constrained at SNR $\gtrsim 10^3$, which in our notation would mean $\sigtot\lesssim 10^{-3}$.  
In \cref{fig:parametric_disk}, we illustrate how well parametric models of stellar surfaces may be measured at $\sigtot\lesssim 10^{-3}$.  The radial intensity profile $I(r)$ is somewhat constrained at $\sigtot = 10^{-3}$ (left panel), however significant degeneracies between the overall size and limb darkening parameters persist at this SNR.  But at even slightly higher SNR ($\sigtot=3\times10^{-4}$, middle panel) the radial profile is tightly constrained, and at even higher SNR ($\sigtot = 10^{-4}$, right panel) uncertainty in $I(r)$ is effectively negligible.  
These SNR thresholds required for tightly constraining $I(r)$ may be understood by inspecting the bottom panels of \cref{fig:parametric_disk}, which show the size of the signal of $I(r)$ relative to a uniform disk, in comparison to the size of the errors \sigV\ corresponding to each different noise level \sigtot.  We cannot constrain $I(r)$ when the binned \sigV\ exceeds the difference in \Vsq\ between the limb-darkened profile and the uniform disk, and conversely $I(r)$ is tightly constrained when the binned \sigV\ is much smaller than the difference between the full profile and uniform disk.  The same statement is plotted in a different way in \cref{fig:profile_fidelity} below.  However, note that even at SNR levels where $I(r)$ is essentially perfectly constrained, full (non-parametric) image reconstruction is not necessarily possible.  We next turn to studying the SNR requirements of image reconstruction.

\subsection{Image reconstruction with noise}

\begin{figure}
\centering
\includegraphics[width=\textwidth]{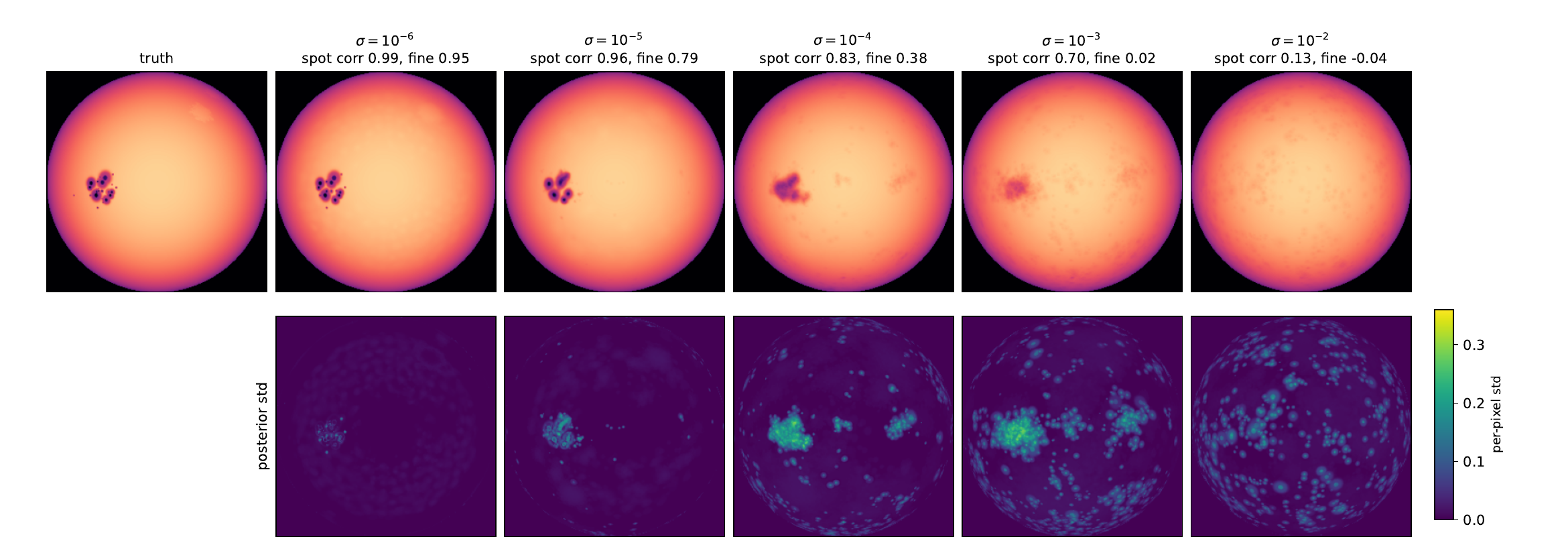}\\[4pt]
\includegraphics[width=0.85\textwidth]{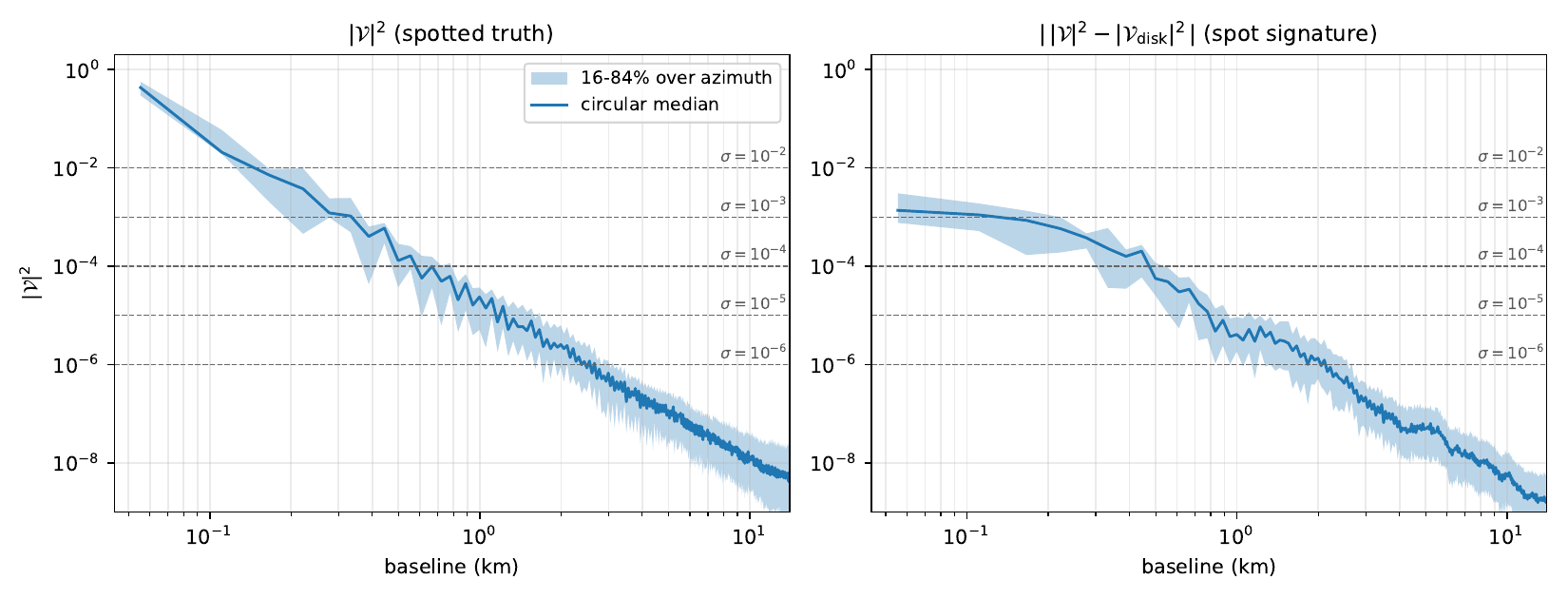}
\caption{Information recovered as a function of SNR. \emph{Top row:} the true
surface (left) and the mean of 32 posterior samples for
noise $\sigV$ indicated in the labels above each panel.
Labels also give the spot correlation of
the mean, and of its high-pass component (``fine''). Structure is recovered scale by scale: the
complex's envelope arrives robustly by $\sigV=10^{-4}$, its fine structure only at
the lowest noise, and at $\sigV\ge10^{-3}$ the mean expresses at most a
preference for the correct location. \emph{Middle row:} the per-pixel
standard deviation of the same samples. At high SNR, variance is confined to
the edges of securely detected spots; at low SNR, every feature in the mean
carries comparable sample-to-sample variance, suggesting that structure there is a preference, not a detection. Individual samples near the detection threshold
can also place partial ghost images at the $180^\circ$-rotated position (the
two-fold degeneracy of phase-less data; \S\ref{sec:metrics}), as can be seen at $\sigV=10^{-3}-10^{-4}$.
\emph{Bottom:} the circularly averaged \Vsq\ of the truth (left; band =
16--84\% range over azimuth) and its spot signature
$\bigl||\mathcal{V}|^2-|\mathcal{V}_{\rm disk}|^2\bigr|$ (right), with the
panel noise levels overlaid. Features at a given baseline are recoverable
roughly where their signature exceeds $\sigma$, which is why the envelope
(short baselines) appears first and the fine structure ($\sim$3~km baselines)
requires the deepest data.
}
\label{fig:retrieval_example}
\end{figure}

Classical phase retrieval \citep{Fienup1982} requires nearly noiseless measurements of \Vsq, with extremely dense baseline coverage (see Appendix \ref{app:hio}).  \Cref{fig:hio,fig:ood,fig:sdo} in \S\ref{sec:validation} show image reconstruction and phase retrieval in the negligible-noise, dense-coverage regime.  For actual astronomical observations, noise is finite and baseline coverage may be limited, so it is important to determine when non-parametric image reconstruction is feasible, and what information we can expect to extract from a given set of observations.

\Cref{fig:retrieval_example} shows an example of the kind of information that can be extracted at various SNR levels.  The behavior shown in the figure may be understood by recalling that small-scale features generally appear at long baselines where \Vsq\ is small, meaning that  low noise levels are required for those small-scale features to appear in reconstructions.  Larger-scale features appear at shorter baselines where \Vsq\ is relatively larger, allowing them to be recovered even at relatively higher noise levels.  In \cref{fig:retrieval_example}, the truth image contains a somewhat large-scale complex of spots, which themselves contain smaller-scale, fine features.  At the lowest SNR (highest noise) levels, we cannot reliably recover any of this structure, as shown in the right-most panel.  As we increase the SNR, posterior samples still exhibit large scatter, but the average of those samples indicates a preference for spots to be placed in proximity to the correct location.  This shows that we first recover large-scale (short-baseline) information.  As the SNR continues to grow, we fill in more of the small-scale, higher-resolution detail, until at the very highest SNR we achieve nearly perfect reconstruction.  Note that this same behaviour was already seen in \cref{fig:ood}.  We caution that this simple picture can be complicated at moderate SNR, right at the threshold where phase retrieval marginally becomes possible, by the two-fold rotational degeneracy of phase-less data (\S\ref{sec:metrics}): we occasionally find marginally detected structures with ghost reflections located on the opposite side of the disk ($\pi$ rotated).  These ghost rotations disappear at larger SNR, so they seem to occur only for marginal detections.  Interestingly, we see similar behavior in classical phase retrieval solutions at much higher SNR (see Appendix \ref{app:hio}).

The takeaway message here is that image reconstruction becomes possible at high SNR, but the exact SNR threshold for reliable reconstruction depends on the scale of the features being recovered.  Since small-scale features are seen only on long baselines, this means that besides SNR, baseline coverage is also important for image reconstruction. 

\subsection{Baseline coverage and array geometry}

\begin{figure*}
\includegraphics[width=\textwidth]{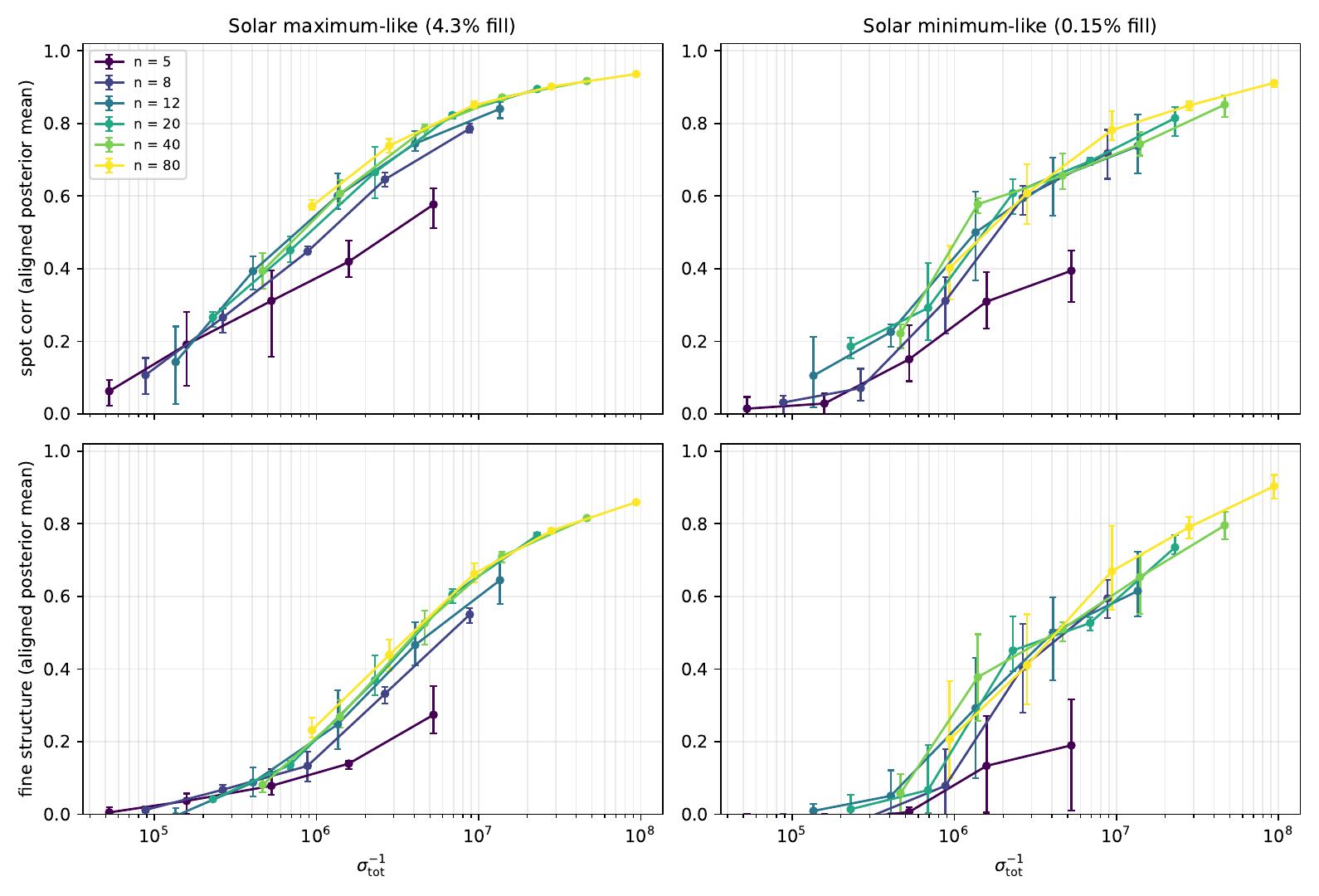}
\caption{Reconstruction fidelity vs.\ array noise \sigtot.  Arrays consist of $n$ telescopes with 
random layout plus compact core, including Earth-rotation synthesis, with $\Bmax=4$\,km.  The x-axis shows the combined array noise, plotted as $\sigtot^{-1}$ so that the SNR increases towards the right.  The top row shows \texttt{spot\_corr} of the posterior mean, while the bottom row shows fidelity of fine-structure reconstruction.  The left panels show reconstruction of a more active star, and right panels are for a more quiet star.  Points and error bars show mean and min/max over independent realizations of the array layout, noise, and sampler seed. Note that for each realization we average over only 4 posterior samples. The small number of samples systematically underestimates the fidelity of the average by an amount which is typically smaller than the plotted uncertainties.  For
$n\gtrsim10$ the curves for different $n$ overlap each other across two decades of noise amplitude, showing how fidelity depends only on the aggregate
SNR, not on details of the array or individual baselines' \sigV.  The exception is at very sparse arrays that are coverage-limited, as illustrated by $n=5$ array. Its reconstruction saturates at low fidelity regardless of SNR, and its fine-structure recovery stays near zero, showing that  unmeasured Fourier cells cannot always be
traded for photons.}
\label{fig:ntel}
\end{figure*}

\begin{figure*}
\includegraphics[width=\textwidth]{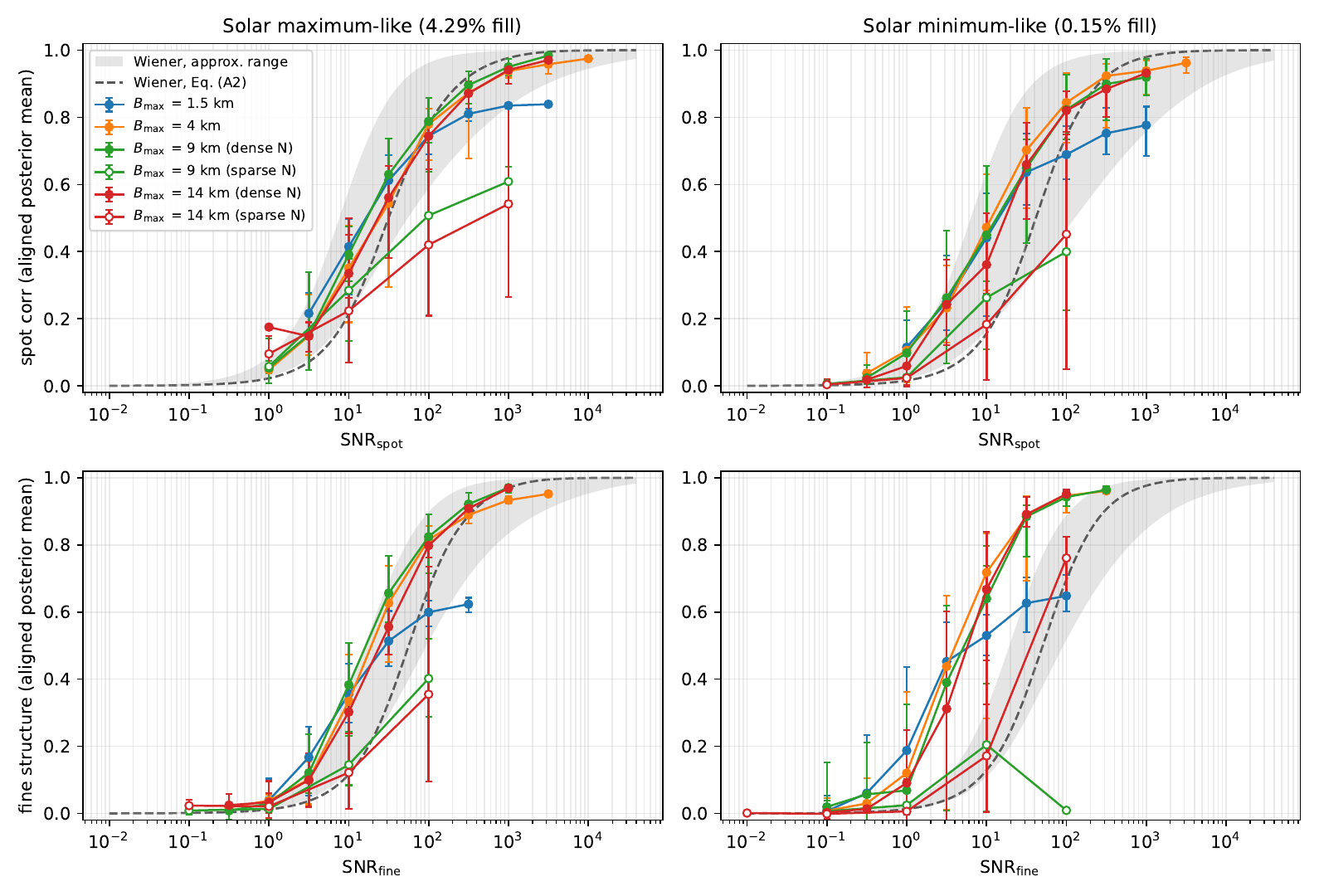}
\caption{Reconstruction fidelity vs.\ signal-to-noise ratio.  Compared to \cref{fig:ntel}, instead of noise level here the x-axis shows the SNR as defined in \cref{eq:snrmatched}. We show results both for spot correlations (top row) and fine structure (bottom row).  Different colors show arrays with different \Bmax.  Error bars correspond to the scatter within bins, and each measurement was an average of only 4 posterior samples. The small number of samples systematically underestimates the fidelity of the average by an amount which is typically smaller than the plotted uncertainties.  The close agreement of the different curves shows that reconstruction fidelity depends mainly on this SNR, and at fixed SNR is insensitive to details of the array configuration or individual baseline noise \sigV.  Two exceptions are also apparent.  First, short-baseline arrays that under-resolve the features of interest cannot reconstruct those features, even with arbitrarily large SNR.  This can be seen in the plateau of the $\Bmax=1.5\;$km curves.  Second, a minimal coverage density is required for robust phase retrieval, as can be seen in the ``sparse $N$'' points with open symbols, which also fall far below the other curves even at the same SNR.  The sparse points have $N<3000$ at $\Bmax=9\,$km, and $N<10^4$ at $\Bmax=14\,$km. These exceptions show that additional photons cannot completely compensate for missing information in the Fourier plane.  For comparison, the shaded band shows the range of Gaussian (Wiener) predictions of \cref{eq:wiener_corr} in Appendix~\ref{app:Wiener}.  In some cases (e.g., the low-activity star) the reconstruction surpasses this Gaussian prediction, presumably by making use of non-Gaussianity of the signal.}
\label{fig:snr_signature}
\end{figure*}

\begin{figure*}
\includegraphics[width=\textwidth]{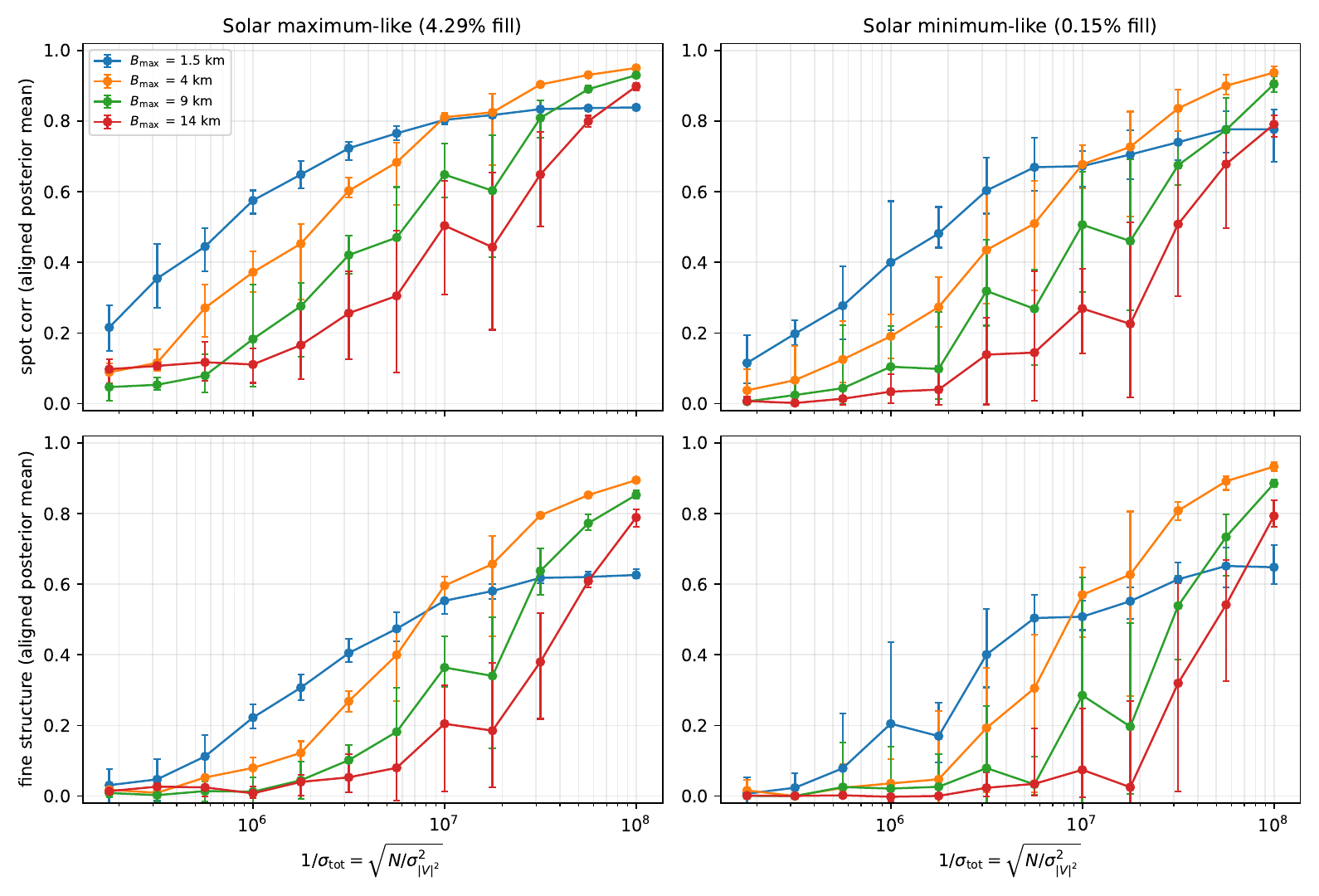}
\caption{Same as \cref{fig:snr_signature}, but replotted with $\sigtot^{-1}$ on the x-axis instead of SNR.  Unlike at fixed SNR, at fixed array noise we see that reconstruction fidelity does depend on baseline length, because the signal lives on certain baselines.  Arrays that set their sizes to match the angular scales of the features of interest can reconstruct those features at higher fidelity than similar arrays that have either larger or smaller sizes.}
\label{fig:snr}
\end{figure*}

\begin{figure}
\centering
\includegraphics[width=0.65\textwidth]{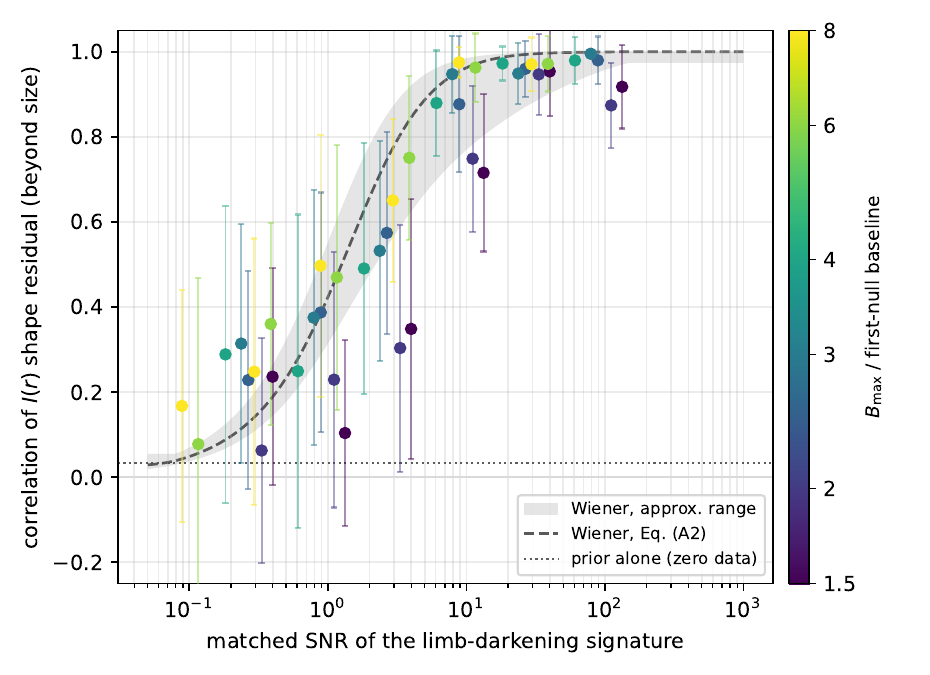}
\caption{The fidelity-SNR relation for the parametric reconstruction 
of stellar intensity profile $I(r)$.  Points show the 5-parameter
(size + limb-darkening) fits of \cref{fig:parametric_disk} over a grid of
baseline reach $B_{\rm max}$ (colors) and noise \sigtot\ (mean $\pm$ scatter
over 24 noise realizations). The fidelity metric isolates the profile
\emph{shape}: each reconstructed profile has its own best-fit uniform disk
subtracted, projecting out the overall size, and the result is correlated
against the same residual of the truth. The x-axis is the matched SNR of the
limb-darkening signature,
${\rm SNR}^2=\sum_{\bm B}\bigl[\bigl(\Vsq(\bm B)-\Vsq_{\rm ud}(\bm
B)\bigr)/\sigV\bigr]^2$, where $\Vsq_{\rm ud}$ is the best-fitting uniform
disk with the size re-optimized.  The dotted
line marks the measured zero-data floor (the prior and model family alone).
The shaded band spans the Gaussian (Wiener) readings of \cref{eq:wiener_corr}
(cf.\ \cref{fig:snr_signature}), evaluated per coverage with this signature's
shell spectrum, and the dashed curve is the power-proportional reading.
As in \cref{fig:snr_signature}, arrays too short to sample the signature
plateau below unit fidelity: the $B_{\rm max}=1.5\times$-null points (darkest)
saturate around $0.8$--$0.9$ while $B_{\rm max}\geq4\times$ converge to unity.}
\label{fig:profile_fidelity}
\end{figure}

The previous examples of reconstructions have used all possible independent Fourier-space measurements of \Vsq, up to the Nyquist frequency given by the image resolution.  Real interferometers do not have densely packed telescopes sampling all possible baselines, so next we examine the impact of sparse coverage of the Fourier plane. \Cref{fig:ntel} shows image reconstruction fidelity for two example types of stars, roughly corresponding to surface activity for the Sun near the maximum of its 11-year cycle (left panels) or near the minimum (right panels).  Different curves show reconstruction fidelity, as quantified by the metrics discussed in \S\ref{sec:metrics}, for arrays with various numbers of telescopes.  The example arrays place $n$ telescopes randomly with maximum baseline lengths given by $\Bmax=4\;$km, including a subset at short baselines forming a compact core.  We do not give much detail about the array geometry, because as can be seen from the Figure, the exact details of the array geometry turn out to be relatively unimportant, once a minimum threshold number of telescopes is exceeded.  

For the different random telescope arrays, we perform image reconstruction at various noise levels $\sigma$ and number of telescopes $n$.  \cref{fig:ntel} shows results for $\Bmax=4\;$km, and \cref{fig:snr_signature} shows similar results for other \Bmax.  The key result shown in \cref{fig:ntel,fig:snr_signature} is that, except at the very lowest number of telescopes, image reconstruction fidelity appears to depend on $n$, $\sigma$, and array geometry specifically via the signal to noise ratio of the features of interest in the image.  We can start to detect features in the image when their matched-filter SNR exceeds $\sim10$, and reconstruction of those features becomes robust (high-fidelity) when this SNR reaches $\sim10^2$ (\cref{fig:snr_signature}).  Here the matched-filter SNR is the significance of the features' own signature in the data, computed over the measured baselines, and weighted by the spatial scales that the fidelity metric measures,
\begin{equation}
{\rm SNR}_x^2 \;=\; \sum_{\bm{B}} \left[  \frac{W_{x}(\bm{B}) \left(\Vsq(\bm{B}) - \Vsq_{\rm disk}(\bm{B})\right)}{\sigV} \right]^2,
\label{eq:snrmatched}
\end{equation}
where the sum runs over baseline vectors $\bm{B}$, and $W_x(\bm{B})$ is a spatial filter to pick out the Fourier modes of the signal: for ${\rm SNR}_{\rm fine}$, $W_{\rm fine}$ is the same high-pass filter used to define the fine structure metric (a $12$-pixel uniform box subtracted from the identity), and for ${\rm SNR}_{\rm spot}$, $W_{\rm spot}$ is the same construction at 48 pixels, which removes only the whole-disk-scale cells but leaves untouched spots.  Results are insensitive to varying the size of spot filter, as long as the size is large compared to spot sizes, and small compared to the size of the stellar disk.  Note that this expression does not use the optimal Wiener filter, but instead uses a much simpler high-pass filter.  The insensitivity to filter scale, and the results shown in \cref{fig:snr_signature}, suggest that this sub-optimality does not significantly degrade the results.

As \cref{fig:snr_signature} shows, the fidelity with which features are reconstructed depends on the SNR of those features in approximately the manner expected for an optimal filter.  This is most evident in the left panels, showing fidelity for a star with relatively high surface activity, in comparison with the Wiener filter prediction (grey dashed line) for this star's surface power spectrum; see Appendix \ref{app:Wiener} for more details. 
Roughly speaking, features are detected when their SNR $\gtrsim 10$ and reconstructed reliably when their SNR $\gtrsim 100$.  
An interesting counter-example is shown in the right panels of the Figure, corresponding to a star with a relatively quiet surface.  Here, reconstruction fidelity \emph{outperforms} the Wiener filter.  Presumably, this is because reconstruction exploits the non-Gaussianity of the spots and faculae, which becomes even more pronounced when those surface features are sparse, whereas the Wiener filter uses only information in the power spectrum.   
But for our purposes here, the main takeaway from this Figure is that image reconstruction depends on the details of the telescope array in an extremely simple manner, only through the SNR.  At fixed SNR, the details of array geometry, number of baselines, telescope size, etc., appear relatively unimportant.

\cref{fig:snr_signature} also highlights two important exceptions to the above statement.  First, in order to reconstruct image features with an array, that array must contain baselines long enough to resolve those features.  This is illustrated by the blue curves and points, corresponding to $\Bmax=1.5$ km.  Unlike the other curves, reconstruction fidelity plateaus for $\Bmax=1.5$ km, even as we increase the total SNR.  This makes sense: more precisely measuring short-baseline modes that under-resolve surface features cannot replace missing information on scales that do resolve those features.  However, once sufficiently long baselines are present, then the exact value of \Bmax\ becomes unimportant, at fixed SNR.  The other exception is array sparsity, illustrated by the points with $n=5$ in \cref{fig:ntel}, and also illustrated by the open symbols in \cref{fig:snr_signature} (labeled ``sparse $N$'').  These correspond to arrays with low numbers of independent baselines (accounting for Earth rotation), $N<3000$ for $\Bmax=9$ km and $N<10^4$ for $\Bmax=14$ km. As these curves indicate, overly sparse arrays do not reach the same fidelity as denser arrays at the same SNR.  Once again, better measuring a tiny subset of Fourier modes cannot completely replace the information missing in the unmeasured modes.  However, above a feature-dependent coverage threshold, increased coverage density does not further improve reconstruction at fixed SNR, as demonstrated by the rest of the curves, which span orders of magnitude in $N$, the number of independent baselines.  
The practical lesson is that for arrays capable of resolving the features of interest and that exceed a very minimal density of baseline coverage, improving reconstruction fidelity only requires collecting more photons, irrespective of whether those photons are collected using a large number of small telescopes or a small (but not too small) number of large telescopes.

Note that holding fixed the signal to noise ratio is not the same as holding fixed the noise, because the signal depends on baseline.  This is seen in \cref{fig:snr}, which re-plots the same results as in \cref{fig:snr_signature}, but using noise instead of SNR as the x-axis.  In this figure, we find quite different reconstruction fidelity as a function of \Bmax, unlike the earlier figure, simply because the signal is found on certain baselines.  This illustrates the well-known property of interferometric imaging that tuning array size (characterized here by \Bmax) to match the features that we seek to recover can improve our recovery of those features.

Lastly, the behavior shown in \cref{fig:snr_signature} is not unique to spot recovery, but instead arises more generally.  For example, in \cref{fig:parametric_disk} we showed reconstruction of the circularly averaged $I(r)$ profile using parametric models.  The reconstruction fidelity of $I(r)$ using parametric fits depends on the profile SNR in a qualitatively similar manner as spot fidelity depends on spot SNR, as illustrated in \cref{fig:profile_fidelity}.  In detail, profile fidelity grows more steeply with profile SNR (compared to spot reconstruction) because the parametric model has fewer degrees of freedom than our non-parametric maps of spots and faculae.

\section{Conclusions}
\label{sec:discussion}

In this paper, we have studied Bayesian image reconstruction using phaseless visibility data of the kind provided by intensity interferometry.  
We showed that faithful reconstructions of observed sources are possible from phase-less data, recovering complex
structure in the source's appearance.  The Bayesian approach makes use of physical priors on source structure, and one key takeaway is that 
high-fidelity image reconstruction is possible even when those priors are mis-specified or inaccurate.  
It may be surprising that so little prior knowledge suffices to recover Fourier phases from a vast sea of possible values, but classical 
phase retrieval already demonstrates that the very weakest prior knowledge is sufficient to determine Fourier phases.  The Bayesian
approach offers significant benefits over classical iterative phase retrieval, giving reconstructions that are far more robust to measurement 
noise and sparse sampling of the Fourier plane.

We showed that score-based diffusion models provide an extremely flexible and general method to construct Bayesian priors for intensity interferometry. 
We constructed a Bayesian reconstruction pipeline relying on diffusion priors, having in mind stellar intensity interferometry as a specific example.  
However, the method can be applied to any other source observed by intensity interferometers, as long as a vaguely plausible training set can be
constructed for those sources.

Using our reconstruction pipeline, we studied the information that can be extracted from intensity interferometers.  Our main finding is shown in \cref{fig:snr_signature} (and also in \cref{fig:ntel,fig:profile_fidelity}). We find that reconstruction of features in an image depends almost entirely on the signal to noise ratio alone of those specific features, be they small-scale spots or the large-scale mean profile $I(r)$ of the source.  \emph{At fixed SNR}, reconstruction is almost insensitive to details of the telescope array like configuration, collecting area, baseline coverage, etc., although \cref{fig:snr_signature} shows important exceptions to that conclusion.  We also find that reconstruction fidelity depends on SNR (very!) approximately following the predictions of Gaussian Wiener filtering.  When images contain non-Gaussian structure, however, that structure can be recovered with fidelity surpassing the Gaussian prediction. 

This result means that we can straightforwardly predict what aspects of a source's appearance could be reconstructed by different telescope arrays, just by knowing the array's noise.  Ballpark forecasts can even be constructed without needing the full reconstruction pipeline, since the Gaussian estimate needs only the power spectra of the source and of the noise.  Our results show that at SNR levels achievable in the near future by next-generation intensity interferometers, high-fidelity non-parametric image reconstruction may be performed for bright, nearby stars.
In future work we will study what science we can learn about astrophysical sources using intensity interferometry observations.

\section*{Software and data}
JAX \citep{jax2018}, Flax \citep{flax2020github}, Optax \citep{deepmind2020jax}, PyTorch\citep{paszke2019pytorch}, 
healpy \citep{Zonca2019healpy,Gorski2005healpix}, spotter \citep{spotter},
NumPy \citep{Harris2020numpy}, SciPy \citep{Virtanen2020scipy}, Matplotlib
\citep{Hunter2007matplotlib}, Astropy \citep{astropy2013,astropy2018,astropy2022},
and SunPy/drms \citep{sunpy2020,Glogowski2019drms} for the SDO fetch. 
SDO/HMI data courtesy of NASA/SDO and the HMI science team.

\section*{Acknowledgments}
We thank Marios Galanis and Yashar Hezaveh for helpful discussions, and thank Lizhong Zhang and George N. Wong for sharing GRMHD simulations on black holes.  ND thanks the Aspen
Center for Physics for hospitality during the final completion of this work.
BD acknowledges support from the Ambrose Monell Foundation, the Corning Glass Works Foundation Fellowship Fund, and the Institute for Advanced study. MB was supported in part by the UK Science and Technology Facilities Council under grant agreement ST/Y0021561/1.
Research at Perimeter Institute is supported in part by the Government of Canada through the Department of Innovation, Science and Economic Development Canada and by the Province of Ontario through the Ministry of Economic Development, Job Creation and Trade. This work was enabled in part by resources provided by Compute Ontario and the Digital Research Alliance of Canada.
This work made substantial use of Claude (Anthropic), accessed through Claude Code:
AI assistance was used to implement and validate the analysis software, run and
monitor the numerical experiments, prepare figures, and assist in drafting this
manuscript. All results were specified, reviewed, and verified by the authors, who
take full responsibility for the content of this paper.

\bibliography{references}

\appendix

\section{Expected reconstruction fidelity}
\label{app:Wiener}

\cref{fig:snr_signature} shows several examples in which reconstruction fidelity depends principally on overall signal-to-noise ratio.  In this appendix, we provide a simple analytic estimate of what fidelity can be expected as a function of SNR, using only the power spectrum of the expected signal.  We also briefly discuss the impact of prior misspecification on the SNR required to achieve a given fidelity.

Reconstruction fidelity can be anticipated from the standard theory of linear
estimation \citep{VanTrees1968,Kay1998detection}. Consider first the quantity
that is actually measured: in each Fourier cell $k$ of the $(u,v)$ plane, the
windowed \Vsq\ signature of \cref{eq:snrmatched},
$s_k = W\,(\Vsq-\Vsq_{\rm disk})_k$, is observed as $d_k = s_k + n_k$ with
independent noise of variance $\sigV^2$. For a signal of prior variance
$p_k^2$, the minimum-variance linear estimate is the Wiener filter
$\hat s_k = w_k d_k$, with gain and residual variance
\begin{equation}
w_k = \frac{p_k^2}{p_k^2+\sigma^2} = \frac{\rho_k^2}{1+\rho_k^2},
\qquad
\big\langle(\hat s_k-s_k)^2\big\rangle = w_k\,\sigma^2,
\qquad
\rho_k^2 \equiv \frac{p_k^2}{\sigma^2}.
\label{eq:wiener_gain}
\end{equation}
Well-measured modes ($\rho_k\gg1$) pass through unchanged ($w_k\to1$), while
noise-dominated modes ($\rho_k\ll1$) are shrunk toward the prior mean
($w_k\to0$) rather than passing their full noise into the map. When the prior
is \emph{matched} to the signal, $p_k^2=\langle s_k^2\rangle$, the per-mode SNR
is $\rho_k = W|\Vsq-\Vsq_{\rm disk}|_k/\sigV$, and the total
$\rho^2\equiv\sum_k\rho_k^2$ is exactly the squared matched-filter SNR of
\cref{eq:snrmatched}. The fidelity of any linear estimate assembled from these
gains follows immediately: if the quantity being scored has power $q_k$ in mode
$k$, the correlation between estimate and truth is the signal-power-weighted
mean of the gain,
\begin{equation}
{\rm corr}^2 = \frac{\sum_k w_k\,q_k}{\sum_k q_k},
\qquad w_k = \frac{\rho_k^2}{1+\rho_k^2}.
\label{eq:wiener_corr}
\end{equation}

What \cref{eq:wiener_corr} predicts depends on what is being estimated, and
here the phase-less data force a choice, because the quantity we measure (the
\Vsq\ signature) is not the quantity the fidelity metrics score (the image).
We consider three options. (i)~With $q_k = p_k^2$, \cref{eq:wiener_corr}
gives the fidelity with which the \Vsq-signature map itself is
\emph{denoised}. For this conventional choice the measurement model above is
exact --- but it is a statement about the data, not the image: converting even
perfectly denoised moduli into an image is precisely the phase-retrieval step,
about which it says nothing. (ii)~To address the image, take
$q_k = W^2|R_k|^2$, the windowed power spectrum of the residual-image modes
($R_k$ = the Fourier transform of the image minus the unspotted disk), and
suppose each mode is recovered at the SNR its \Vsq\ signature carries ---
the same $\rho_k$ as above. Because a mode's \Vsq\ signature scales roughly as
the \emph{square} of its amplitude, weak modes are doubly suppressed under
this assignment, making it a conservative choice for image fidelity.
(iii)~Alternatively, distribute the same total SNR over the image modes in
proportion to their power, $\rho_k^2 = \rho^2\, q_k/\sum_j q_j$ --- the
assignment that would hold if the modes were measured linearly with white
noise. None of the three follows rigorously from the phase-less likelihood
(that would require the Fisher information of \Vsq\ data about each image
mode, which is beyond our scope here). Options (i) and (ii) bracket the
possibilities generously from above and below, and we show them as the shaded
band in \cref{fig:snr_signature}; the intermediate option (iii) is drawn as
the dashed curve. All three share the same qualitative shape: because a
localized feature is broadband in the Fourier plane, its power spans a wide
range of $\rho_k$, so the modes switch on gradually as $\rho$ grows and the
transition is broad --- matching the reconstructions --- rather than the sharp
step a single characteristic SNR would give. Conversely, a signature carried
by a few smooth modes --- the limb-darkening profile of
\cref{fig:profile_fidelity} --- switches on over a narrower range, at matched
SNR of order unity. We stress that \cref{eq:snrmatched} weights the signature
by a fixed high-pass filter, not by the optimal (Wiener) gain, so
\cref{eq:wiener_corr} would not hold exactly even for a Gaussian signal; the
close agreement in \cref{fig:snr_signature} shows this simple filter is
already a good proxy for the metric's scales.

Crucially, every choice in \cref{eq:wiener_corr} uses only \emph{power
spectra} ($p_k^2$, $q_k$): each is the optimal linear reconstruction for a
stationary Gaussian random field with that spectrum.  Our signals, however,
can be significantly non-Gaussian since they are composed of small numbers of
spots and faculae that themselves are spatially clustered.  We can expect them
to have significant bispectra and other higher $N$-point functions encoding
the phase correlations that produce these sparse compact features, and
\cref{eq:wiener_corr} is blind to that non-Gaussianity.
\Cref{eq:wiener_corr} is therefore a Gaussian benchmark: the best
fidelity a power-spectrum prior can reach within the linear idealization,
with the width of the band in \cref{fig:snr_signature} expressing the
ambiguity of applying that idealization to phase-less data. We would not
expect reconstruction fidelity exceeding the entire band to be produced
by power-spectrum information alone.

In contrast, the diffusion prior is not Gaussian. 
It has learned the distribution of spot maps, e.g.\ that realizations 
contain sparse, compact features.  This is precisely the
higher-order structure that \cref{eq:wiener_corr} discards, which is why
our diffusion model-based reconstruction can
exceed the Gaussian benchmark, recovering sharp features from partial data
(super-resolution). The excess grows with the non-Gaussianity of 
the surface features, as illustrated in \cref{fig:snr_signature}.

A mismatched prior degrades reconstruction fidelity in various ways. 
A Gaussian prior with the wrong power spectrum, $p_k^2\neq s_k^2$, has correlation
\begin{equation}
{\rm corr}\;\approx\;\frac{\sum_k w_k s_k^2}
{\sqrt{\big(\sum_k w_k^2(s_k^2+\sigma^2)\big)\big(\sum_k s_k^2\big)}}\,,
\qquad w_k=\frac{p_k^2}{p_k^2+\sigma^2}\,,
\label{eq:wiener_general}
\end{equation}
below the matched value of \cref{eq:wiener_corr}. Note that the estimate is still
$w_k d_k$, with mean $w_k s_k$, meaning that the map is the truth scaled toward the prior mean
(zero). In this case, the lowered fidelity is \emph{loss}, either over-smoothing toward blank
(over-confident prior) or excess noise (under-confident prior), but is not due to invented
structure. In contrast, a non-Gaussian prior mismatched in shape behaves differently,
because its ``mean'' is not simply $w_k d_k$, but instead is a low-dimensional manifold of spot-like
maps.  When the truth is representable but atypical, like a feature in the tails of
the prior, then the misspecification bias fades gracefully, diminishing with (modest) extra SNR.
But if the truth lies outside the prior's support entirely, then the prior 
projects the estimate onto its manifold, inventing prior-typical structure 
that the data do not demand. Therefore, the same non-Gaussian power that super-resolves 
for a matched prior now super-resolves the wrong structure for a mismatched prior,  
capping fidelity until the SNR overwhelms the prior (\S\ref{sec:validation}). 
Nongaussianity of the prior is thus double-edged, the source of both super-resolution 
and hallucination, while the Gaussian benchmark has neither aspect.

\begin{figure}
\centering
\includegraphics[width=\textwidth]{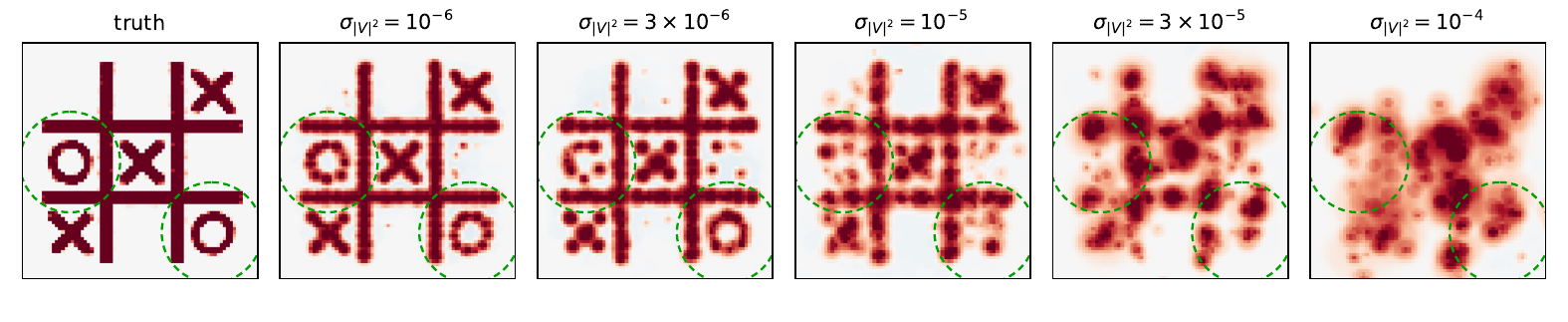}\\
\includegraphics[width=0.58\textwidth]{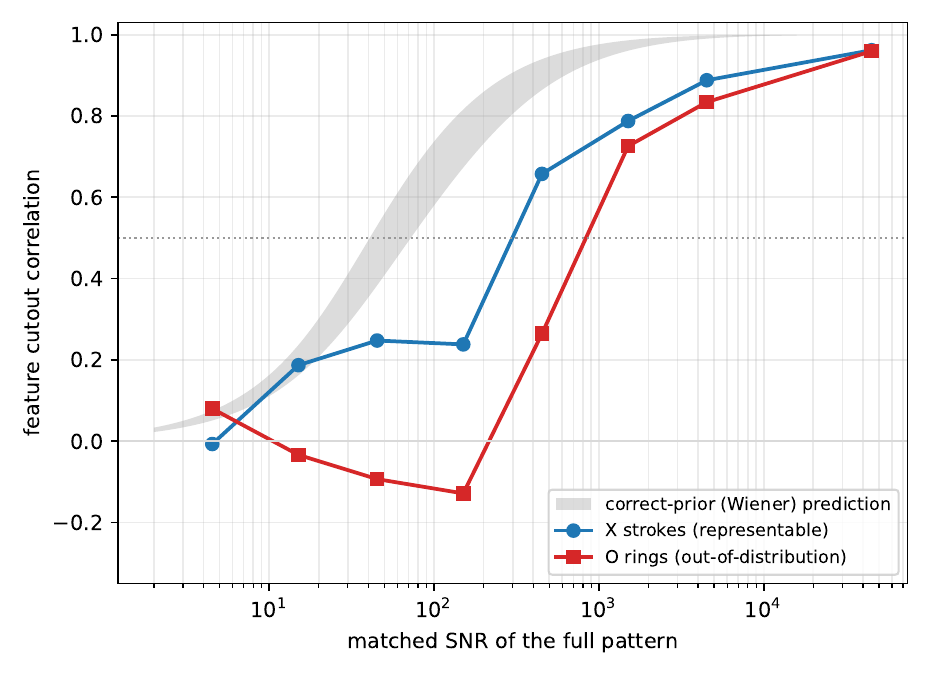}
\caption{Differential effect of prior mis-specification, from the tic-tac-toe
example of \cref{fig:ood}. \emph{Top:} truth and posterior-mean reconstructions
at decreasing SNR (full Fourier coverage); dashed circles mark the O rings. The
prior fills the ring holes at low SNR, placing a compact spot inside
the hole at intermediate SNR, and the data re-open them at high SNR.
\emph{Bottom:} per-feature cutout correlation versus the matched SNR of the
full pattern, averaged over the three X strokes (blue) and two O rings (red).
The gray band is the corresponding prediction of \cref{eq:wiener_corr} for a
correct (matched-Gaussian) prior, spanning the whole-pattern and per-class
signature spectra. The representable X strokes lag the correct-prior band by a
factor of $\sim\!5$--$7$ in SNR; the out-of-distribution rings by
$\sim\!12$--$20$, and their correlation is transiently negative where
the prior draws a spot in the ring's hole.}
\label{fig:tictactoe}
\end{figure}

These regimes are borne out by the out-of-distribution tests of
\S\ref{sec:validation}. Mildly atypical features are recovered with modest extra
SNR, simply moving the observation along the curve of \cref{fig:snr_signature},
whereas qualitatively unrepresented morphology requires several times the
in-distribution SNR before the data override the prior. 
The tic-tac-toe example of \cref{fig:ood} contains both cases side by side: its
grid and X strokes are representable morphology (assembled from spot-like
features), while its O's are rings, a topology further outside the prior's support.
Because both classes sit in the same image, reconstructed in the same sampler
run from the same noise realizations, comparing them isolates the effect of
shape alone.  \Cref{fig:tictactoe} makes the comparison quantitative, comparing 
fidelity of individual X's and O's.  We score each feature by the mean-subtracted 
correlation between the true and mean-reconstructed pixel
values within the smallest square box bounding it, plotted against the matched
SNR of the full pattern, and compare against the prediction of
\cref{eq:wiener_corr} for a hypothetical correct prior, i.e.\ the Wiener
fidelity evaluated with the tic-tac-toe's own power spectrum.  For both X's and O's, 
fidelity lags the Wiener band, unlike the behaviour seen in \cref{fig:snr_signature},  
illustrating the price of the mis-specified prior. Given the sample size (and
the crudeness of comparing a cutout statistic with a global Gaussian benchmark) we
regard these as order-of-magnitude illustrations rather than precision
measurements, but even such ballpark numbers convey how much reconstruction
fidelity costs under different kinds of prior mis-specification, largest near
the detection threshold and shrinking as the data override the prior.

\section{Sampler validation details}
\label{app:sampler}

All reconstructions in this work use
the annealed reverse-SDE sampler of \S\ref{sec:sampling}. Diffusion-based
posterior sampling is an active area, and a number of other approaches have
been discussed in the literature, including diffusion posterior sampling
\citep{ChungDPS2023}, variational approaches \citep{Mardani2023reddiff},
plug-and-play Langevin MCMC \citep{Laumont2022pnpula}, microcanonical Langevin
MCMC \citep{Robnik2023mclmc}, and twisted sequential Monte Carlo
\citep{Wu2023tds}; \citet{Daras2024survey} survey the field.

\begin{figure}
\centering
\includegraphics[width=\textwidth]{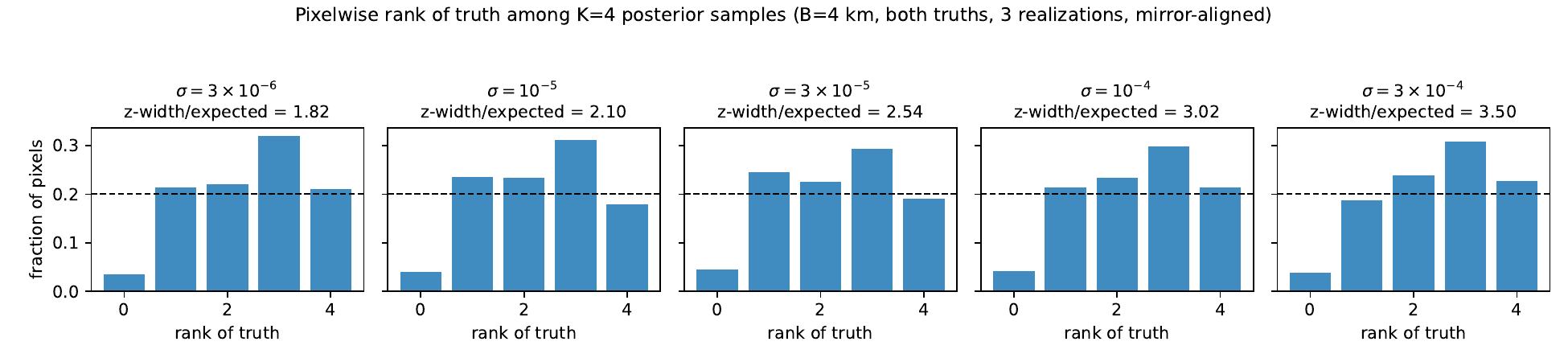}
\caption{Rank-based calibration diagnostic from the survey ensembles
(\S\ref{sec:results}; $\Bmax=4$\,km, both truths, three realizations per cell).
Each panel pools, over visible-disk pixels, the rank of the true pixel value
among the $K=4$ mirror-aligned posterior samples; a calibrated posterior is
uniform (dashed line). The bulk is approximately uniform at every noise level;
the depleted lowest rank reflects a mild tendency of samples to over-generate
faint dark structure, and the pooled $z$-score widths (quoted above each panel,
relative to the small-$K$ Student-$t$ expectation) reveal heavy tails from
sub-threshold features that the ensemble confidently lacks.}
\label{fig:calibration}
\end{figure}

We perform two different posterior miscalibration tests on our sampler. First, the rank test 
of \cref{fig:calibration} examines the posterior one pixel at a time. For
each visible-disk pixel it ranks the true value among the $K$ samples 
(aligning to remove the $\pi$-rotation degeneracy). A calibrated marginal gives 
a uniform rank on average 
\citep[the rank-histogram test of simulation-based calibration;][]{Talts2018sbc}.
This test is a 1-point statistic that provides no information about
correlations between pixels, i.e. a posterior can be properly calibrated at every pixel
separately and still place the structure wrongly on the disk.
We perform this test on examples from 
\S\ref{sec:results} using $K=4$ samples. The bulk of the
histogram is uniform to within $\sim50\%$ at every noise level, so the
ensembles neither collapse onto a point nor scatter grossly. Two deviations are
informative. The lowest rank is depleted $\sim5\times$: the samples slightly
over-generate faint dark structure, the same mild filling bias noted in
\S\ref{sec:results}. The pooled $z$-scores are $1.8$--$3.5\times$ wider than
the small-$K$ expectation and grow towards high $\sigma$. The posterior is
approximately correct about features above the detection threshold and
overconfident about the absence of features below it.

\begin{figure}
\centering
\includegraphics[width=\textwidth]{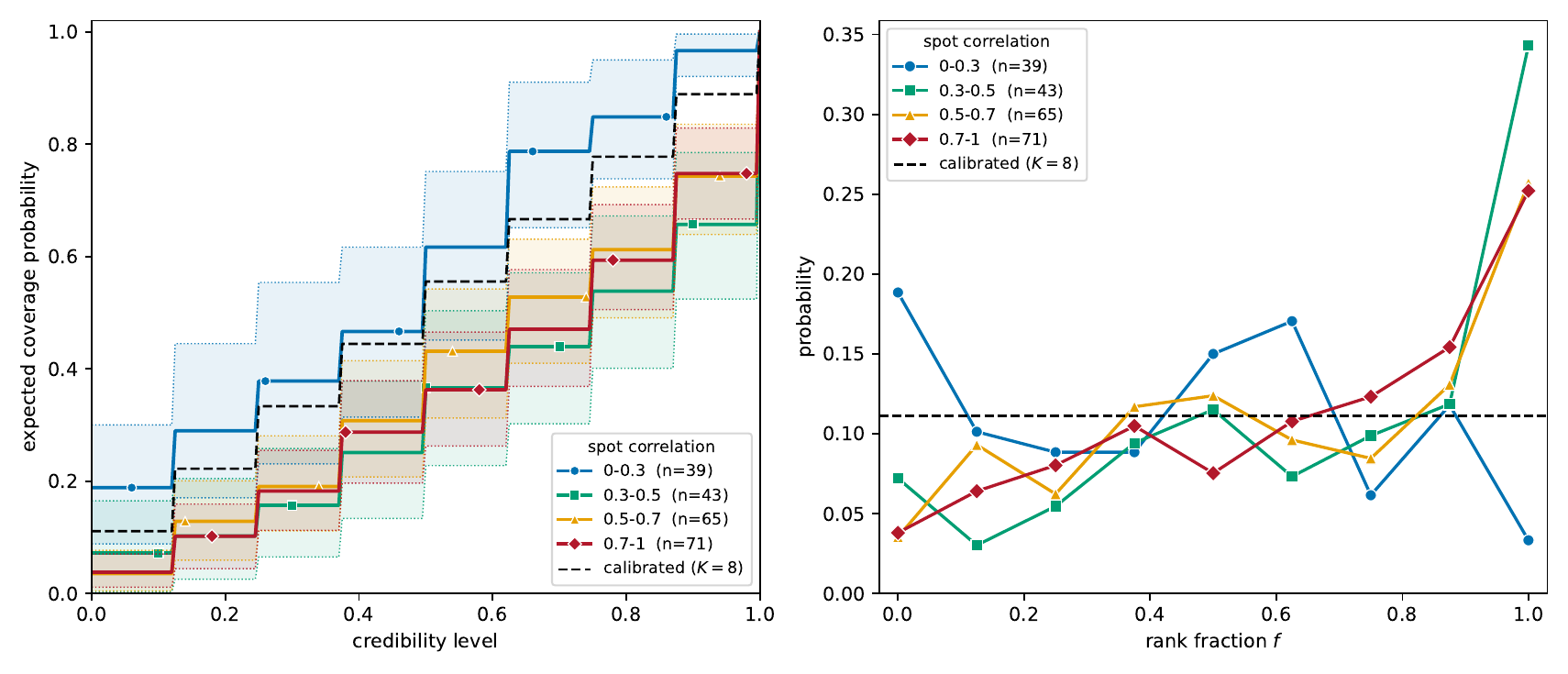}
\caption{TARP expected-coverage test of the full reconstruction pipeline.
Truth images are drawn from the training-set generator, and distances are measured between the
observed disks. Simulations spanning a range of truth images and noise levels
are pooled and binned by spot correlation of the mean reconstruction averaged
over $K=8$ posterior samples each. \emph{Left:} coverage against credibility, with bootstrap 95\%
bands and the calibrated $K=8$ staircase. We find that at low fidelity 
(${\tt spot\_corr} < 0.3$) the posterior is underconfident, 
but otherwise the posterior is  overconfident. \emph{Right:} the
same data plotted as differential probability, instead of cumulative. 
The mild miscalibration seen in the left panel arises from a pile-up at $f=1$ ---
$P(f{=}1)=0.03$, $0.34$, $0.26$, $0.25$ in the four bins against the
calibrated $0.11$, while $P(f{=}0)$ runs the other way. This corresponds to 
a one-sided bias toward under-recovered feature power, not a symmetric 
underestimate of the posterior width.}
\label{fig:tarp}
\end{figure}

A complementary, fully simulation-based check
is the TARP expected-coverage test \citep{Lemos2023tarp}, the validation
applied to the same problem class by \citet{Dia2025iris}.  Unlike the per-pixel 
test described above, the TARP test of \cref{fig:tarp} examines the joint
distribution instead. It reduces each map to a single distance from a reference 
map and asks whether the truth is exchangeable with the samples in that distance. 
TARP measures how much structure the posterior recovers and is nearly blind to
where that structure lies.  To perform this test, an ensemble of truth maps is 
drawn from the training-set generator, and each is reconstructed from its
own noisy \Vsq\ realization ($K=8$ posterior samples, using the sampler
configuration of \S\ref{sec:sampling}). The truth's distance to a random
reference point is ranked against the $K$ samples' distances to the same
reference. If the posterior is calibrated, the truth is statistically
exchangeable with the samples, so the rank fraction $f$ is uniform and its CDF
(the expected coverage probability, ECP) follows the finite-$K$ staircase.
Reference points are the truths of the other simulations (exact generator
draws), and distances are measured between intrinsic deficit maps weighted by
$\sqrt{\mu}\,{\rm LD}(\mu)$, so that the squared distance equals the squared
difference of the corresponding observed disks (limb darkening is held fixed
and known within this test). Note that because the truths come from the
generator while the reconstruction uses the learned diffusion prior, this test
probes the combined prior$+$sampler pipeline against the true data-generating
distribution, a point we return to below.

We test for posterior miscalibration at various levels of reconstruction fidelity.
In all cases we observe evidence of miscalibration, typically mild.  
The sign of the miscalibration reverses across the figure. At low fidelity 
(${\tt spot\_corr} < 0.3$), the posterior is over-dispersed, meaning that the
truth falls closer to the reference than every sample more often than it should.
For example, $P(f{=}0)=0.19$ against the calibrated $1/(K{+}1)=0.11$. 
For larger fidelity, 
the posterior is overconfident, and the deviation is one-sided. The probability
of the top rank climbs to $P(f{=}1)=0.34$, $0.26$ and $0.25$ in the three higher
bins while $P(f{=}0)$ falls to $0.03$--$0.04$. A posterior that was merely too
narrow would inflate both tails together. Because the distance to a reference
map of disjoint support is dominated by each map's own weighted feature power,
a pile-up at $f=1$ instead means the truth carries more feature power than all
$K$ samples. The posterior under-recovers feature power by about one unit of
its own scatter, which in these regimes is small.
Diffusion posterior sampling exhibits a similar loss of sample diversity \citep{ChungDPS2023, Xu2025dpsmap}, though the underlying mechanism may differ from ours.

\section{Prior training details} \label{app:training}

In \S\ref{sec:diffusionprior}, we briefly described the training of our diffusion
prior.  Below, we discuss several aspects of the prior training in more detail.

\begin{figure*}
\includegraphics[width=\textwidth]{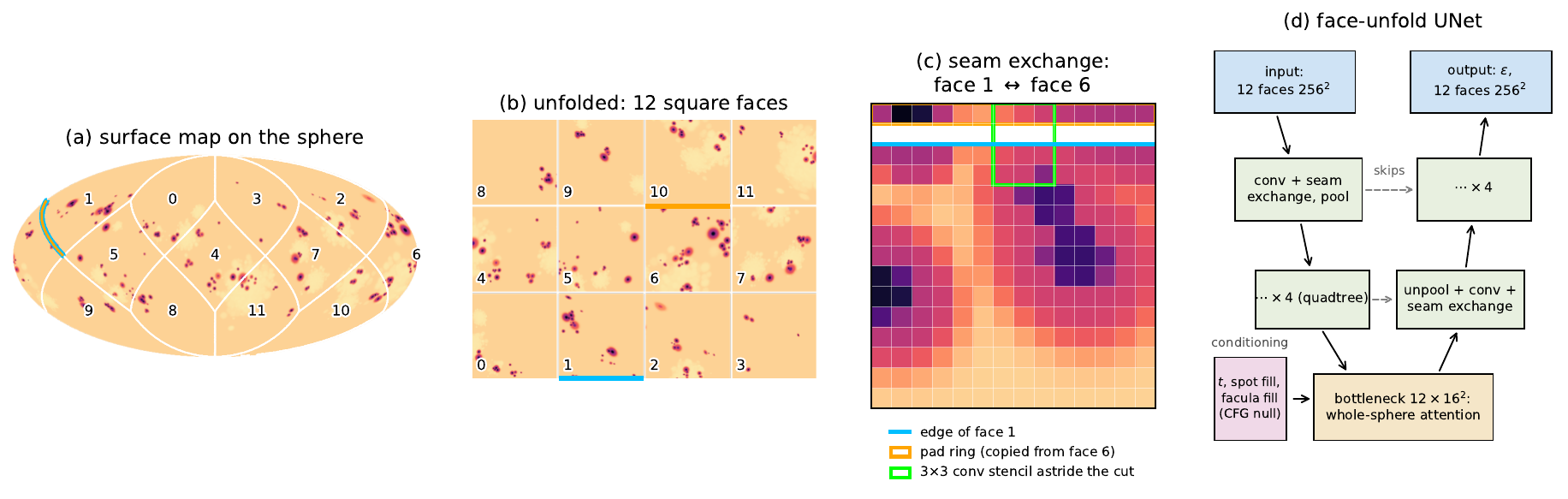}
\caption{The HEALPix face-unfold architecture of the diffusion prior. \emph{(a)} A surface map drawn from the training generator (an active star, for visibility), with the boundaries of the 12 equal-area HEALPix base tiles overlaid and numbered: tiles 0--3 ring the north pole, 4--7 the equator, 8--11 the south pole. \emph{(b)} The same map unfolded: in the NESTED pixelization each tile is a perfect $N_{\rm side}\times N_{\rm side}$ square grid, so the sphere becomes a stack of 12 ordinary images on which standard 2D convolutions can operate. The cuts, however, separate structure that is contiguous on the sphere. \emph{(c)} The seam exchange that repairs this: before every $3\times3$ convolution each face is padded with a one-pixel ring copied from its geometric neighbors on the sphere.  Shown is the edge of face 1 (blue), whose pad ring (orange, drawn lifted off the face) is gathered from the matching edge of face 6; the same two edges are marked in panel (b), and the single physical seam they share is marked on the sphere in panel (a).  Faces 1 and 6 are not adjacent in the unfolded stack --- one seam on the sphere becomes two distant edges after cutting --- which is why the exchange must be derived from the true sphere neighbor relations rather than simple array padding.  A convolution stencil astride the cut (green) therefore sees the uncut sphere, and features crossing tile boundaries are processed identically to features in a tile interior. \emph{(d)} The resulting denoising UNet: per-face convolutions with seam exchange, four exact quadtree poolings ($N_{\rm side}=256\to16$), a bottleneck self-attention layer over all $12\times16^2$ coarse tokens (providing a whole-sphere receptive field for global statistics such as the total spot coverage), and skip connections; the diffusion time $t$ and the spot- and facula-filling factors enter as conditioning inputs.}
\label{fig:arch}
\end{figure*}

Our denoising network is a U-Net
\citep{Ronneberger2015unet}, the encoder--decoder convolutional
architecture with skip connections that is the standard backbone for diffusion
models \citep{HoDDPM2020}. The network operates on the 12 unfolded HEALPix base faces
of the NESTED pixelization (\cref{fig:arch}). Convolutions are applied
per-face with the cross-face seam exchange derived geometrically from the true sphere neighbor relations.  
The prior's training set consists of $5\times10^4$
intrinsic deficit maps $\delta(\hat n)$ from the generator of \S\ref{sec:generator}, stored natively at HEALPix $N_{\rm side}=256$ ($786{,}432$ pixels per map). Training used Adam \citep{KingmaBa2015adam} with a cosine learning-rate decay
($2\times10^{-4}\to5\times10^{-5}$), gradient
checkpointing, and effective batch sizes ranging from 8 to 128. 
All sampling uses an exponential moving average (EMA) of the
weights (decay $0.999$), the standard stabilizer for score-based models
\citep{SongErmon2020improved}. 

Our trained DDPM prior exhibits clear signs of under-training.  The distributions of
spot coverage and facular coverage of samples drawn from this prior do not match
the corresponding distributions of the images that make up the prior's training
set.  For instance, $\sim 8\%$ of prior samples have spot coverage fractions larger
than the maximum allowed by the training distribution, and a similar fraction of 
prior samples have facular coverage fractions larger than allowed by the training
distribution.

We considered a few training variants to mitigate this defect 
(loss reweighting \citep{Hang2023minsnr}, conditioning
and guidance \citep{HoSalimans2022cfg}, effective batch sizes up to 128),
before deciding to use this under-trained prior as is, since this under-training
does not appear to produce significant changes in our main results.  
For example, under-training should cause the fidelity-SNR relations shown in 
\cref{fig:ntel,fig:snr_signature,fig:snr} to be somewhat biased due to prior 
mis-specification.  The truth images used in those examples were drawn
from the same generator that produced the training set, and not sampled from the
learned prior, which has a somewhat different distribution.  For this 
reason, the fidelity at a given SNR shown in those examples should be biased low.
However, we can estimate the size of this SNR penalty by repeating our fidelity 
estimates using truth images sampled from the learned prior, rather than the 
generator used to construct the training set.  Based on (admittedly small) 
experiments, we do not find any significant evidence for an SNR penalty due
to prior under-training.

It is worth noting that this artifact appears to be peculiar to the 
DDPM prior training recipe that we happened to use as our first training 
attempt in this study, and is not a general deficiency of diffusion models. 
For example, prior samples from a model trained with the EDM recipe
\citep{Karras2022edm} do not exhibit the large excesses of spots or faculae
that some of our DDPM samples show.

\section{Comparison to classical iterative phase retrieval}
\label{app:hio}

We implemented Fienup's hybrid input-output (HIO)
algorithm \citep{Fienup1982} and applied it to the same phase-less data that our
sampler faces. The implementation uses hard Fourier-modulus
projection on measured cells, object-domain support (the exact stellar disk) and
non-negativity, $\beta=0.9$ feedback with periodic error-reduction blocks, the
twin-breaking asymmetric support reduction of \citet{FienupWackerman1986}
(required here: a centered stellar disk is precisely the centrosymmetric-support
case for which HIO notoriously stagnates in object+twin mixtures), and multiple
random restarts with the returned solution selected by the data residual,
not by the real-space true image. 
We validated exact recovery (correlation 1.000, residual at
machine precision) on random compact objects, where HIO is known to work well, before applying it to stellar images. 
For these tests, we used images with $128^2$ pixels (results at $256^2$ are consistent).  For each reconstruction, we use 16 random initializations, apply $3\times10^4$ iterations each, and use the best-fitting (in \Vsq\ space) realization.

On noiseless, complete $\Vsq$ data, HIO reaches
$\texttt{spot\_corr}=0.72$ for a solar-minimum-like truth (two small active
regions) and $0.55$ for the solar-maximum-like truth (46 regions), against
$\approx1.00$ for the diffusion-prior posterior on equivalent data
(\cref{fig:hio}). Although HIO has notoriously slow convergence, 
this low fidelity appears to be stagnation, since
the failure mode is structurally specific. Decomposing images into parts symmetric
and antisymmetric under the $180^{\circ}$ image rotation (the exact symmetry of
phase-less data), HIO recovers the symmetric component nearly perfectly,
with correlation $0.998$ ($0.984$) at $99\%$ ($98\%$) amplitude for the two
truths, while the antisymmetric component is recovered at
correlation $-0.44$ ($-0.14$). The HIO solutions are mixtures of the true image
and its $\pi$ rotation
($\approx0.27{:}0.73$ and $0.42{:}0.56$). This split has a simple
origin: the symmetric image component has a real Fourier transform, so its
``phase retrieval'' degenerates to sign retrieval, which support and positivity
constraints solve. In contrast, the antisymmetric component carries the genuinely lost
continuous phase, and the classical constraint set happily admits any mixture of the 
two degenerate solutions. Our sampler's individual posterior draws, by contrast, 
each commit to a pure version of an image or its $\pi$-rotation (\S\ref{sec:metrics}), 
because a blend has little support under the surface prior.

With noise, the contrast between HIO and posterior sampling sharpens dramatically. Adding Gaussian noise of width
$\sigV=10^{-7}$ to $\Vsq$, which is smaller in amplitude than every feature 
signature in the data, 
collapses HIO to $\texttt{spot\_corr}=0.14$ ($0.11$ at $10^{-6}$, $0.03$ at
$10^{-5}$, as shown in \cref{fig:hio}), while the posterior is
unaffected at $\sigV=10^{-7}$, and has $\texttt{spot\_corr}=0.68$ at 
$\sigV=10^{-5}$.  Noise-tolerant classical variants of phase
retrieval (relaxed or tolerance projections) degrade less abruptly but supply 
neither a statistical noise model nor any uncertainty quantification.

Incomplete Fourier coverage is equally destructive.  Randomly removing even
$10\%$ of the cells halves the recovered fidelity ($\texttt{spot\_corr}=0.50$), and by
$70\%$ coverage the reconstruction collapses outright ($0.14$).  In this regime, HIO reconstructions have stripe artifacts filling the disk, a common
failure mode described in \citealt{FienupWackerman1986}. Prior-based
reconstruction in this work routinely operates at $50\%$ random coverage with
no material penalty, and our earlier coverage studies recovered structure at
random fractions down to $\sim6\%$.
\cref{fig:hio} (left column) makes the contrast explicit with a matched head-to-head
at $15\%$ coverage, where no algorithm relying on the data alone can succeed. 
At this coverage, HIO returns structureless noise
($\texttt{spot\_corr}=0.00$) with a data residual better than its
successful full-coverage runs (the sparse constraints are satisfied exactly, by
garbage), while the diffusion-prior posterior mean on the same measured cells
recovers the map at $\texttt{spot\_corr}=0.97$. Classical iteration thus
fails a factor $\gtrsim3$ above the information-counting threshold while the
prior-based posterior succeeds well below it. 

\end{document}